\documentclass[11pt,a4paper]{article}
\usepackage{jheppub}
\usepackage[utf8]{inputenc}
\usepackage{amsmath,amssymb}
\usepackage{ytableau}
\ytableausetup{centertableaux,boxsize=1.em}
\usepackage{multirow}

\usepackage{pifont}

\usepackage{color}
\usepackage{graphicx}
\usepackage{amsfonts}
\usepackage{bm}
\usepackage{epsfig}
\usepackage{eufrak}
\usepackage{tikz}
\usepackage{tikz-cd}
\usepackage{ctable}

\usepackage{tikz} 
\usetikzlibrary{decorations.markings} 
\usetikzlibrary{decorations.text} 
\usetikzlibrary{positioning} 
\usetikzlibrary{backgrounds} 
\usetikzlibrary{fit} 
\usetikzlibrary{calc} 
\usetikzlibrary{shapes} 

\tikzset{dot/.style={draw,circle,inner sep=.7pt,fill,node
    distance=1cm}} 
\tikzset{dot1/.style={draw,circle,inner sep=.7pt,fill}} 
\tikzset{triangle/.style={draw,regular polygon, regular polygon
    sides=3}} 
\tikzset{->-/.style={decoration={
  markings,
  mark=at position .5 with {\arrow{>}}},postaction={decorate}}} 
\tikzset{-<-/.style={decoration={ 
  markings,
  mark=at position .5 with {\arrow{<}}},postaction={decorate}}}

\newcommand\dd{{\rm d}}

\newcommand\be{\begin{equation}}
\newcommand\ee{\end{equation}}
\newcommand\bea{\begin{eqnarray}}
\newcommand\eea{\end{eqnarray}}

\begin{document}

\begin{titlepage}
\renewcommand{\thefootnote}{\fnsymbol{footnote}}

\vspace*{1.0cm}

\begin{center}
{\textbf{\huge A Note on Circle Compactification \vskip0.3cm of Tensile Ambitwistor String }}
\end{center}
\vspace{1.0cm}

\centerline{
\textsc{\large Kanghoon Lee,} $^{a}$%
\footnote{kanghoon.lee1@gmail.com} \hskip0.5cm
\textsc{\large J. A.  Rosabal} $^{a}$%
\footnote{j.alejandro.rosabal@gmail.com}
}

\vspace{0.6cm}

\begin{center}
${}^a${\it Fields, Gravity \& Strings @ {\rm CTPU}, Institute for Basic Science, \\
70 Yuseong-daero 1689-gil, Daejeon 34047, \rm KOREA}
\end{center}

\vspace{2cm}
\centerline{\bf Abstract}
\begin{centerline}
\noindent
We discuss a number of problems associated with the circle compactification of the bosonic tensile ambitwistor string with the  asymmetric vacuum  choice. By considering the spectrum and  physical state conditions, we show that the circle radius plays a role as a tuning parameter which determines the low energy effective field theory. At the self dual point, we construct the current operators and compute OPEs between them. While the final outcome remains as yet inconclusive, several new results are obtained. Through the current algebra analysis we show that there is a subsector of the massless states where the gauge symmetry is enhanced to $ISO(2)_L\times SU(2)_R$. Using the fact that the one loop partition function is not modular invariant, we show that tensile ambitwistor string partition function is exactly the same as a field theory partition function. Our result proves that despite the existence of  winding modes, which is a typical characteristic of a string, the tensile ambitwistor string behaves as a point particle theory.
\end{centerline}

\thispagestyle{empty}

\end{titlepage}

\setcounter{footnote}{0}

\tableofcontents

\section{Introduction}

Strings, unlike point particles, are extended objects that are able to wind around non-contractible cycles. Their extended structure is responsible for many interesting phenomena that are not evident in a field theory of point particles. For instance, in the case of toroidal compactification, besides the Kaluza-Klein (KK) states associated to compact momenta, there are winding states associated to the number of times that a cycle is wrapped by a string. 

At specific points of the moduli space of the torus some massive states become massless and can give rise to enhanced gauge symmetries. The simplest example is provided by the compactification of the ordinary bosonic string on a circle at the self dual radius $R=\sqrt{\alpha^{\prime}}$. At this point the $U(1)_L\times U(1)_R$ gauge symmetry of the KK massless modes enhances to $SU(2)_L\times SU(2)_R$.

In the last few years there have been several attempts to construct new string theories \cite{Mason:2013sva, Siegel:2015axg,  Huang:2016bdd, Leite:2016fno, Casali:2016atr, Lee:2017utr, Casali:2017mss}. They have tried to explain the worldsheet origin of several apparently unrelated theories, such as HSZ theory \cite{Hohm:2013jaa, Hohm:2016lim} in double field theory or the CHY formula \cite{Cachazo:2013gna, Cachazo:2013hca, Cachazo:2013iea, Cachazo:2014nsa} in scattering amplitudes. Recently Casali and Tourkine clarified \cite{Casali:2016atr} the relation between ambitwistor and tensionless  string theory by means of an asymmetric vacuum choice between the left and right oscillators \cite{Gamboa:1989px,Gamboa:1989zc}. Subsequently  in \cite{Lee:2017utr} a consistent quantization scheme was proposed for the finite string tension case of the ambitwistor string. 

A remarkable common feature of these new string theories that we shall refer to as {\it twisted string theories}\footnote{As in \cite{Casali:2017mss} we shall refer to this kind of theories as {\it twisted strings}. Its possible mathematical foundations could be related with so called {\it twisted cocycle} \cite{Mizera:2017rqa}.} is the finite number of degrees of freedom. When quantizing  twisted string theory on a non-compact flat spacetime  a finite number of degrees of freedom comes out. It is regarded as a theory living at the edge of quantum field theory (QFT) and string theory. For instance, scattering amplitudes of twisted string theory are invariant under $s\leftrightarrow t \leftrightarrow u$ exchange as in conventional string theory, however, at the same time, the partition function of twisted string theory is not modular invariant \cite{Lee:2017utr, Yu:2017bpw} as in QFT.

On the one hand, there are evidences \cite{Adamo:2014wea, Adamo:2013tsa} that suggests  supersymmetric twisted string theories are quantum mechanically consistent. For instance the non-compact flat supersymmetric extension is ghost free  \cite{Mason:2013sva, Huang:2016bdd}. Nevertheless full consistency has not been proven yet.
On the other hand, when twisted string theory is compactified on tori, an infinite number of KK and winding states arises, and the spectrum changes  significantly  \cite{Casali:2017mss}. Interestingly, even though the infinite tower of massive winding excitations reveals stringy features that are not appreciable from the non-compact twisted string, the theory still possesses QFT features. However, the appearance of infinitely many negative norm states, even in the supersymmetric extension of the twisted compact string, makes the full quantum consistency subtle and perhaps impossible to achieve. Regardless of quantum consistency, compactified twisted string theory deserves more attention and a deeper exploration to understand the essence of this new string theory.

In this paper we study bosonic twisted string theory compactified on a circle with an arbitrary radius $R$. We focus on the most distinctive features of string theory that a generic QFT does not have, such as
\begin{itemize}
\item The spectrum: infinitely many degrees of freedom are regarded as a key property of string theory. While QFT contains a finite number of degrees of freedom.
\item Gauge symmetry enhancement: this important stringy effect of circle compactification arises at the self dual radius. For the conventional string the $U(1) \times U(1)$ gauge symmetry is enhanced to $SU(2)\times SU(2)$.

\item The partition function: string theories have modular invariant partition functions. However, in QFT modular invariance is not a quantum requirement. 
\end{itemize}
We investigate these features in twisted string theory and compare with the conventional string theory. 

We show that twisted string is sensitive to the circle radius $R$, and the physical spectrum is drastically altered by changing $R$. We find that there are particular points where the number of massless fields reduce to 4: the massless NSNS fields $g, B $ and $\phi$, and a higher spin field $\bar{a}_{\mu_{1}\mu_{2}\dots \mu_{n}}$. 

The physical spectrum for a generic $R$ has infinitely many massive higher spin states. At the self dual point infinitely  many of these states become massless. By computing the current algebra we find some evidences that the gauge symmetry is enhanced. Even though the full gauge group is not obvious, for a subsector  of the massless states we identify the left gauge group $G_{L} = ISO(2)$, which is the group contraction of $SO(3)$ .

The partition function for  twisted string compactified on a circle is not modular invariant as in the non-compact case. Interestingly enough, even with infinitely many degrees of freedom, the integration over the moduli space can be performed analytically. This partition function properly counts  the physical degrees of freedom and corresponds to the expected QFT partition function. Twisted string theory, even containing infinitely many degrees of freedom, keeps some features of QFT.

In section \ref{sec2} we briefly review the quantization of string theory with the asymmetric vacuum on flat non-compact spacetime. In section \ref{sec:bosonic} we develop the quantization of twisted bosonic string theory compactified on a circle.  We pay special attention to the spectrum and the points where the number of massless fields is enhanced. In section \ref{sec4}, with the aim of identifying the enhanced symmetry, we compute the OPE and current algebra. We present  one case in which the enhanced algebra can be easily identified, at least for the zero mode. In section \ref{partition_func} we compute the partition function. It turns out that  the partition function, like its non-compact counterpart, is not modular invariant. We compare  the physical spectrum derived directly from the physical conditions with the counting of the degrees of freedom from the partition function, and we find perfect agreement.  Conclusions are presented in section \ref{conclusions}.

\section{Review of closed string theory with asymmetric vacuum}\label{sec2}

In this section we shall review the quantization of bosonic closed string theory with the asymmetric vacuum choice \cite{Lee:2017utr}. The classical theory is the same as ordinary string theory \cite{Green:1987sp, Polchinski:1998rq}. For example the mode expansion is given by 
\bea\label{t_mode_exp}
X_L^\mu (\bar z) &=& X_{0 L }^\mu - i {\alpha' \over 2} P_L^\mu \log \bar z + \sqrt{ \alpha' \over 2} \sum_{n \ne 0} {i \over n} \overline \alpha_n^\mu \bar z^{-n}\,,
\nonumber \\
X_R^\mu (z) &=& X^\mu_{0 R} - i {\alpha' \over 2} P_R^\mu \log z + \sqrt{\alpha' \over 2} \sum_{n \ne 0} {i \over n} \alpha_n^\mu z^{-n}\ .
\eea
Note that the left/right zero modes are treated as two independent operators even for the non-compact Minkowskian target spacetime. However  the eigenvalues of the operators $ P_L^\mu$ and $ P_R^\mu$   in non-compact spacetime coincide,  $k^{\mu}_{L} = k^{\mu}_{R}=k^{\mu}$.

Quantization assumes the ordinary canonical commutation relation
\be
[X^{\mu}(\tau,\sigma), \Pi^{\nu}(\tau,\sigma^{\prime})]=i\ \eta^{\mu\nu}\delta(\sigma-\sigma^{\prime})\,,
\ee
where $\Pi^{\mu}(\tau, \sigma)$ is the conjugate momenta of $X^{\mu}(\tau, \sigma)$
\begin{equation}
  \Pi^{\mu}(\tau,\sigma) = \frac{1}{2\pi \alpha'} \partial_{\tau}X^{\mu}(\tau,\sigma)\,,
\label{}\end{equation}
and their harmonic modes satisfy the usual commutation relations
\begin{equation}
\begin{aligned}
  &\big[\bar \alpha^\mu_m, \bar \alpha^\nu_{n}] = \delta_{m+n,0}\eta^{\mu \nu}\,, & &\big[\alpha^\mu_m, \alpha^\nu_{n}\big] =m \delta_{m+n,0} \eta^{\mu \nu} \,,
  \\
  &\big[X^{\mu}_{0_{L}}, P^{\nu}_{L}\big] = i \eta^{\mu\nu}\,, & &\big[X^{\mu}_{0_{R}}, P^{\nu}_{R}\big] = i\eta^{\mu\nu}\,.
\end{aligned}\label{}
\end{equation}

In the conventional string theory, the left and right vacuum states are defined symmetrically 
\begin{equation}
  \alpha_{m} |0\rangle_{R} = 0\,,\qquad \bar{\alpha}_{m}|0\rangle_{L} = 0\,, \quad\mathrm{for}\quad\, m>0\,.
\label{}\end{equation}
Twisted strings, on the other hand, are constructed with the asymmetric vacuum 
\begin{equation}
  	\left.\begin{aligned}\alpha_{m} |0\rangle_{R} &= 0\,,\qquad & \bar{\alpha}_{-m}|0\rangle_{L} &= 0\,,
  	\\ P_{R} |0\rangle_{0_{R}} &= 0\,, &\sideset{_{0_{L}}}{}{\mathop{\langle 0|}} P_{L} &= 0\,, \end{aligned}\right.\qquad \text{for}~ m>0\,.
\label{}\end{equation}
This can be written more compactly as
\begin{equation}
\alpha_{m} |0\rangle_{R} = 0\,,\qquad \sideset{_{0_{L}}}{}{\mathop{\langle 0|}}  \bar{\alpha}_{m} = 0 \qquad \text{for}~ m \geq 0\,,
\label{vac1111}
\end{equation}
where $\alpha^\mu_{0} = \sqrt{\alpha' \over 2} P^\mu_{R}$ and $\overline{\alpha}^\mu_{0} = \sqrt{\alpha' \over 2} P^\mu_{L}$. 

In order to have a consistent quantization, this vacuum choice has to be complemented  with the backward time ordering for the left sector
\begin{equation}
\left.\begin{aligned} T_{R}\big[B(z_{j}) A(z_{i}) \big] &= A(z_{i}) B(z_{j})\,,
\\ T_{L}\big[\bar{A}(\bar{z}_{i}) \bar{B}(\bar{z}_{j}) \big] &= \bar{B}(\bar{z}_{j}) \bar{A}(\bar{z}_{i})\,, \end{aligned}\right.\qquad \text{for}~ |z_{i}|>|z_{j}|\,.
\end{equation}
The normal ordering prescription for the oscillators, which is  compatible with the time ordering, is to  pull  all $\alpha_{n}^\mu$ and $\overline \alpha_{-n}^\mu$ with $n > 0$ to the right, for example
\begin{equation}
\mathclose:\alpha_{m} \alpha_{-n} \alpha_{p}\mathclose: = \alpha_{-n} \alpha_{m} \alpha_{p} \quad \mbox{and} \quad \mathclose:\overline{\alpha}_{m}\overline{\alpha}_{-n}\overline{\alpha}_{p}\mathclose: = \overline{\alpha}_{m} \overline{\alpha}_{p} \overline{\alpha}_{-n}\,, \qquad \mbox{for} ~ m,n,p>0\,.
\label{norm_ordering}\end{equation}
Having established the prescription one can compute  the two point correlation function
\begin{equation}
  \big\langle 0\big|T\big[X^{\mu}(z_{i},\bar{z}_{i}) X^{\nu}(z_{j},\bar{z}_{j})\big]\big|0\big\rangle  = -\frac{\alpha^{\prime}}{2}\eta^{\mu\nu}\log\Big(\frac{z_i-z_j}{\bar{z}_i-\bar{z}_j}\Big)\,.
\label{two_point_ft}\end{equation}
Notice that the above prescription and definitions lead to a two point function which is translation invariant as expected.

The physical states $ | \mbox{phys} \rangle$  satisfy the Virasoro constraint
\be  \label{matrixelementsPT}
\langle \text{phys} | T_{ab} | \text{phys} \rangle  = 0\ .
\ee
The zero mode of the Virasoro generator is expressed as
\begin{equation}
\begin{aligned}
L_0 & =\frac{1}{2}\alpha_0^2+ N-1\,,
\\
\overline{L}_0 & = \frac{1}{2}\overline{\alpha}_0^2- \bar{N}+1\,,	
\end{aligned}\label{viraope_usual}
\end{equation}
where $N$ and $\bar{N}$ are the level operators
\begin{equation}
\begin{aligned}
  \hat N &= \sum_{n=1}^{\infty}  :\alpha_{-n}^{\mu}\alpha_{n}^{\nu} \eta_{\mu\nu}:\,,
\\
\hat{\bar{N}} &= -\sum_{n=1}^{\infty} :\bar{\alpha}_{n}^{\mu}\bar{\alpha}^{\nu}_{-n}\eta_{\mu\nu}:\,.
\end{aligned}\label{number_def}
\end{equation}

We assume  that $L_0$ acts to the right, but  because of the presence of $P^\mu_{L}$ in  $\overline{L}_0$,  this  acts to the left. Then, the physical conditions should be imposed on the full matrix elements (\ref{matrixelementsPT}). For instance,  
\be 
\langle \text{phys} | L_0 | \text{phys}\rangle \pm\langle \text{phys} | \overline{L}_0 | \text{phys}\rangle=0
\label{phy0}\ .
\ee 
They imply the level-matching constraint
\be
 N+\bar{N}=2,
\ee
and the mass-shell condition, $p_\mu p^\mu = -M^2$, with
\be
M^2=\frac{4}{\alpha^{\prime}}(N-1)=\frac{4}{\alpha^{\prime}}(-\bar{N}+1)\ .
\ee
Therefore, the level-matching constraint allows only 3 spin-2 fields.

Additionally, the Virasoro constraint puts restrictions on the Hilbert space
\be
\langle \text{phys}| L_m | \text{phys}\rangle=0\,,\qquad\qquad \langle \text{phys}| \overline{L}_m | \text{phys}\rangle=0 \label{phys}\ .
\ee
For this vacuum choice it is sufficient to demand  
\bea \label{phys1}
L_m | \text{phys}\rangle & = & 0 \qquad\qquad m>0\,, \\ \nonumber
\langle \text{phys} |\overline{L}_m  & = & 0 \qquad\qquad m>0 \ .
\eea
The physical string spectrum at the critical dimension $D=26$ is summarized in Table \ref{table3}. In order to have a positive norm gravity sector, the  vacuum state $ | 0\rangle$ must be   a  negative norm state
\be
 \langle 0| 0\rangle<0\ .\label{vacnorm}
\ee 
\begin{table}[ht]
\centering
\begin{tabular}{cccccc}
\toprule
$N$ & $\bar{N}$ & $M^2$ & state & gauge condition & norm \\
\midrule
1 & 1 & 0 & $\epsilon_{\mu\nu}\alpha_{-1}^{\mu}\overline{\alpha}_{+1}^{\nu}|0,k\rangle$ & $k^{\mu}\epsilon_{\mu\nu}=k^{\nu}\epsilon_{\mu\nu}=0$ & +1 \\
\midrule
2 & 0 & $+\frac{4}{\alpha^{\prime}}$  & $ a_{\mu\nu} \alpha_{-1}^{\mu}\alpha_{-1}^{\nu}|0,k\rangle$ & $k^{\mu}a_{\mu\nu}= a_{\mu}{}^{\mu}=0$ & $-1$ \\
\midrule
0 & 2 & $-\frac{4}{\alpha^{\prime}}$ & $\overline{a}_{\mu\nu}\overline{\alpha}_{+1}^{\mu}\overline{\alpha}_{+1}^{\nu}|0,k\rangle$ & $k^{\mu}\overline{a}_{\mu\nu}= \overline{a}_{\mu}{}^{\mu}=0$ & $-1$\\
\bottomrule
\end{tabular}
\caption{Spectrum satisfying the Virasoro conditions.}{\label{table3}}
\end{table}

A crucial difference of the asymmetric vacuum choice with respect to ordinary string theory is that upon quantization, a point particle nature emerges. This distinguishing feature was confirmed in the scattering amplitude and one-loop partition function computations \cite{Lee:2017utr}.


\section{Quantizing closed twisted string theory on $M\times S^1$}\label{sec:bosonic}

Unlike the conventional closed string theory quantization, the mass spectrum of the twisted string on a flat Minkowskian spacetime comprises three spin-2 fields: the massless sector of ordinary string theory  plus two additional  ghost fields. Here we shall consider the canonical quantization of the twisted string compactified on a circle. We adopt the same quantization scheme  as for non-compact twisted strings reviewed  in the previous section. For a brief discussion on the quantization of closed twisted string theory on $M\times S^1$ see \cite{Casali:2017mss}.

\subsection{The spectrum with the asymmetric vacuum} \label{spectrum}
In ordinary string theory compactified on a circle, besides the KK modes, additional winding modes arise. Since the classical configuration of the string does not depend on the choice of the vacuum, the mode expansion of  $X^{\mu}(\tau,\sigma)$ is identical to the mode expansion of ordinary string theory on a circle. In the previous section we treated the left/right zero mode operators, $P^{\mu}_{L/R}$ and $X_{0L/R}^{\mu}$, as two independent operators with the same eigenvalue $k^{\mu}$. In contrast, for the $S^1$ compactification, the physical states depend on $K_L^{A}=(k^{\mu},k^{25}_L)$, $K_R^{A}=(k^{\mu},k^{25}_R)$ and $(x^{\mu}, X^{25}_{0L}, X^{25}_{0R})$, where $A$ is the $SO(1,25)$ index which runs $A = 0,1,2,\dots , 25$. Here, $\mu,\nu,\dots$ indices denote the non-compact directions, and 25th direction denotes the circle direction. 

The level operators are identical to the non-compact case,
\begin{equation}
\begin{aligned}
  \hat N &= \sum_{n=1}^{\infty} \Big( :\alpha_{-n}^{\mu}\alpha_{n}^{\nu} \eta_{\mu\nu}:+ :\alpha_{-n}^{25} \alpha_{n}^{25}:\Big)\,,
\\
\hat{\bar{N}} &= -\sum_{n=1}^{\infty} \Big(:\bar{\alpha}_{n}^{\mu}\bar{\alpha}^{\nu}_{-n}\eta_{\mu\nu}: + :\bar{\alpha}_{n}^{25} \bar{\alpha}_{-n}^{25}:\Big)\,.
\end{aligned}\label{number_def}
\end{equation}
The Virasoro operators can be recast as
\bea\label{viraope}
L_0 & = &  \frac{\alpha'}{4}p^{2}+ \frac{\alpha'}{4}p_{R}^{2}+ \hat N-1\,, \\ \nonumber
\bar{L}_0 & = & \frac{\alpha'}{4}p^{2}+ \frac{\alpha'}{4}p_{L}^{2}-\hat{\bar{N}}+1 \,.
\eea
The physical conditions as in  (\ref{matrixelementsPT})  and (\ref{phy0}) imply the level-matching constraint
\begin{equation}
 N +\bar{N}-n\omega - 2 = 0 \ ,
\label{level_matching}\end{equation}
and the mass-shell condition
\be
M^2= \frac{n^2}{R^2}+ \frac{\omega^2 R^2}{\alpha^{\prime 2}}+ \frac{2}{\alpha^{\prime}}(N-\bar{N})\ .
\label{mass_shell}\ee
A direct consequence of (\ref{level_matching}) is that $N$ and $\bar{N}$ can be any positive integers if $n \omega > 0$. For  $n=\omega=0$  there are only 3 solutions as in the non-compact space. However, if $n\neq 0$ or $\omega \neq 0$, there are infinitely many solutions. T-duality invariance of the spectrum follows from the level-matching condition (\ref{level_matching}) and mass formula (\ref{mass_formula1}) \cite{Casali:2017mss},
\begin{equation}
  n \to w, \qquad w\to n \,, \qquad R \to \frac{\alpha'}{R}\,.
\label{T-dual}\end{equation}

A remarkable feature of the twisted string compactified on $S^1$  is that it contains massless higher-spin states even away from the self dual radius $R =  \frac{\alpha'}{R}=\sqrt{\alpha'}$ \cite{Casali:2017mss}.
Adding and subtracting (\ref{level_matching}) and (\ref{mass_shell}), we have
\begin{equation}
  M^{2} = \Big(\frac{n}{R} - \frac{ w R}{\alpha'}\Big)^{2} -\frac{4}{\alpha'} + \frac{4}{\alpha'} N\,,
\label{mass_formula1}\end{equation}
\begin{equation}
  M^{2} = \Big(\frac{n}{R} + \frac{ w R}{\alpha'}\Big)^{2} +\frac{4}{\alpha'} - \frac{4}{\alpha'} \bar{N}\,.
\label{mass_formula2}\end{equation}
Setting  $M^{2}=0$ and using $R = r \sqrt{\alpha'}$, (\ref{mass_formula1}) and (\ref{mass_formula2}) reduce to
\begin{equation}
\begin{aligned}
  N = 1- \frac{1}{4} \Big(\frac{n}{r}- w r\Big)^{2} \,,
  \\
  \bar{N} = 1 + \frac{1}{4}\Big(\frac{n}{r}+w r\Big)^{2}\,.
\end{aligned}\label{massless}
\end{equation}
The first equation of (\ref{massless}) implies that the massless fields can only have the values $N=0$ or $N=1$. For each value of $N$  we get a relation constraining  the values of $n$ and $w$,
and $\bar{N}$ is determined from the second equation of (\ref{massless}). We summarize the result in table 2. 
\begin{table}[ht]
\centering
\begin{tabular}{lcc}
\toprule
 $N$\hspace{0.2cm} & $\bar{N}$ & $n$, $w$\\
  \midrule
 $0$ & $1 + (w r \pm 1)^{2}$ & $n = w r^{2} \pm 2 r $ \\\midrule
 $1$ & $1 + w^{2} r^{2}$ & $n = w r^{2}$\\
 \bottomrule
\end{tabular}
\caption{Massless spectrum}
\end{table}

Note that the number of massless modes  depends on the value of $r$. It is possible to generate various string theories with different  field content by tuning  the value of $r$. For instance, when $r$ is a rational number, infinitely many higher-spin fields arise. 
Another interesting case is when $r$ is given by the following particular form
\begin{equation}
  r = \sqrt{y}+x\,, \qquad x^{-1}\in \mathbb{Z}\,, \quad y\in \mathbb{Q}^{+}\,,\qquad \text{but}\qquad \sqrt{y} \notin \mathbb{Q}^{+}\,,
\label{}\end{equation}
where $\mathbb{Q}^{+}$ is the set of  positive rational numbers. In this case, the massless condition (\ref{massless}) reduces to
%
%
\begin{itemize}
  \item $N=0$,
\begin{equation}
\begin{aligned}
  x&=\mp\frac{1}{w}\,, \qquad n= \pm x+wy\,, \qquad \bar{N} = 2+w n\,,
\end{aligned}\label{}
\end{equation}
\item $N=1$ 
\begin{equation}
  n=0\,, \qquad w=0\,, \qquad\bar{N} = 1\,.
\label{}\end{equation}
\end{itemize}
Therefore, the $N=1$ sector provides the usual massless NSNS sector spectrum, and the $N=0$ sector gives an additional massless higher-spin field depending on the value of $y$. 

For example, the spectrum for $r=\sqrt{2}\pm1$ contains two  positive norm massless spin three  states $\bar{a}_{\mu\nu\rho}^i$, $i=1,2$,
for $(w,n)=(1,1)$ and $(w,n)=(-1,-1)$
 and the ordinary NSNS massless  states $g_{\mu\nu}\,, B_{\mu\nu}\,, \phi$.  Notice that for $r = \sqrt{y}+x$, unlike at the  points where infinitely many massless fields arise,  the infinite tension limit $\alpha'\to 0$ is under control,  because there is a finite number of massless states.  We can generalize this to arbitrary higher spin fields by tuning  the value $y$ appropriately, and the $\alpha'\to 0$ limit provides a single massless higher spin field together with massless NSNS fields. Therefore, the circle radius $r$ is a tuning parameter which controls the low energy effective field theory.

One of the important issues is the existence of ghost modes. In general the sign of the the norm of each state  is determined by the number of left oscillators. A generic state is constructed as
\begin{equation}
  \bar{\alpha}_{1}^{\bar{N}_{1}} \bar{\alpha}_{2}^{\bar{N}_{2}} \bar{\alpha}_{3}^{\bar{N}_{3}} \cdots \bar{\alpha}_{n}^{\bar{N}_{n}} |0\rangle\,.
\label{phy1}\end{equation}
Because of the definition of the vacuum  (\ref{vac1111}) and (\ref{vacnorm}), if the total number of left moving creation operators, $\bar{N}_1+\bar{N}_{2}+\dots + \bar{N}_{n}$, is even (odd), then the norm is negative (positive).

Let us present the first few massless states  with non-vanishing momentum and winding modes at the self-dual radius, $r=1$. 

\begin{itemize}
\item $N+\bar{N} = 1$ ($n w=-1$)
\begin{equation}
\begin{aligned}
N = 0\,, \quad & \bar{N}=1\,, \quad n= \pm1\,, \quad &&w= \mp1\,, \qquad &\text{positive norm}
\end{aligned}\label{}
\end{equation}
\item $N+\bar{N} = 2$ ($n w = 0$)
\begin{equation}
\begin{aligned}
  &N = 0\,, \quad &\bar{N} = 2\,, \quad &n= \pm2\,,\quad &&w=0\,,\quad &\text{negative norm}
  \\
  &N = 0\,, \quad &\bar{N} = 2\,, \quad &n= 0\,,\quad &&w=\pm2\,,\quad &\text{negative norm}
\end{aligned}\label{NbN2}
\end{equation}
  \item $N+\bar{N} = 3$ ($nw=1$)
\begin{equation}
\begin{aligned}
  &N = 1\,, \quad &\bar{N} = 2\,, \quad &n= \pm1\,,\quad &&w=\pm1\,,\quad &\text{negative norm}
\end{aligned}\label{NbN3}
\end{equation}
  \item $N+\bar{N} = 5$ ($nw=3$)
\begin{equation}
\begin{aligned}
  &N = 0\,, \quad &\bar{N} = 5\,, \quad &n= \pm3\,,\quad &&w=\pm1\,,\quad &\text{positive norm}
 \\
 &N = 0\,, \quad &\bar{N} = 5\,, \quad &n= \pm1\,,\quad &&w=\pm3\,,\quad &\text{positive norm}
\end{aligned}\label{}
\end{equation}
  \item $N+\bar{N} = 6$ ($nw=4$)
\begin{equation}
\begin{aligned}
  &N = 1\,, \quad &\bar{N} = 5\,, \quad &n= \pm2\,,\quad &&w=\pm2\,,\quad &\text{positive norm} \ . 
\end{aligned}\label{}
\end{equation}
\end{itemize}
In general the massless states are given by
\begin{equation}
\begin{aligned}
&\text{For}~ N=0\,,
\\
  &\quad|0,\bar{N},k^{\mu},k_{L},k_{R}\rangle = \bar{a}_{B_{1},B_{2},\cdots, B_{n}} \big(\bar{\alpha}_{1}^{B_{1}}\big)^{\bar{N}_{1}} \big(\bar{\alpha}_{2}^{B_{2}}\big)^{\bar{N}_{2}} \cdots \big(\bar{\alpha}_{n}^{B_{n}}\big)^{\bar{N}_{n}} |0,0,k^{\mu},k_{L},k_{R}\rangle\,,
\\
&\text{for}~N=1\,,
\\
  &\quad |1,\bar{N},k^{\mu},k_{L},k_{R}\rangle =\epsilon_{A,B_{1},B_{2},\dots,B_{n}} \alpha_{-1}^A \big(\bar{\alpha}_{1}^{B_{1}}\big)^{\bar{N}_{1}} \big(\bar{\alpha}_{2}^{B_{2}}\big)^{\bar{N}_{2}} \cdots \big(\bar{\alpha}_{n}^{B_{n}}\big)^{\bar{N}_{n}} |0,0,k^{\mu},k_{L},k_{R}\rangle\,,
\end{aligned}\label{genMasslessStates}
\end{equation}
where $\bar{a}_{B_{1},B_{2},\cdots, B_{n}}$ and $\epsilon_{A,B_{1},B_{2},\dots,B_{n}}$ are polarization tensors,
and $|0,0,k^{\mu},k_{L},k_{R}\rangle$ is the ground state satisfying 
\begin{equation}
\begin{aligned}
  p^{\mu} |0,0,k^{\mu},k_{L},k_{R}\rangle &= k^{\mu}|0,0,k^{\mu},k_{L},k_{R}\rangle\,,
  \\
  \alpha^{25}_{0} |0,0,k^{\mu},k_{L},k_{R}\rangle &= k_{R}|0,0,k^{\mu},k_{L},k_{R}\rangle\,,
  \\
  \bar{\alpha}^{25}_{0} |0,0,k^{\mu},k_{L},k_{R}\rangle &= k_{L} |0,0,k^{\mu},k_{L},k_{R}\rangle\,.
\end{aligned}\label{}
\end{equation}

\subsection{Physical state conditions}

Since the physical states should be invariant under the worldsheet diffeomorphism, we have to demand the physical state conditions on the Hilbert space
\be
\langle \text{phys}| L_m | \text{phys}\rangle=0\,,\qquad\qquad \langle \text{phys}| \bar{L}_m | \text{phys}\rangle=0 \label{phys}\ .
\ee
In usual string theory it is enough to demand  $L_m | \text{phys}\rangle=0$ and $\bar{L}_m | \text{phys}\rangle=0$ for $m>0$. Here, due to the  different vacuum choice, we impose the following conditions, which are compatible with (\ref{phys}),
\bea \label{phys1}
L_m | \text{phys}\rangle & = & 0 \qquad\qquad m>0\,, \\ \nonumber
\langle \text{phys} |\bar{L}_m  & = & 0 \qquad\qquad m>0 \ .
\eea
The asymmetric vacuum choice affects only the overall sign of the physical state condition. Thus, the physical state condition for twisted strings can be regarded  as in ordinary string theory. Constructing physical  higher spin states has been studied  in the literature \cite{Sasaki:1985py, Manes:1988gz}. Here we list the tensor representations of the first few physical states:
\begin{itemize}
  \item $N$, $\bar{N} = 1$
\begin{equation}
  \begin{ytableau}
 *(gray){}
\end{ytableau}
\label{}\end{equation}
\item $N$, $\bar{N}=2$ 
\begin{equation}
  \begin{ytableau}
 {} & {}
\end{ytableau}
\label{}\end{equation}
\item $N$, $\bar{N}=3$
\begin{equation}
\begin{aligned}
  \begin{ytableau} {} & {} & {} \end{ytableau}\,,
  \quad \begin{ytableau} {}\\{}\end{ytableau}
\end{aligned}\label{}
\end{equation}
\item $N$, $\bar{N}=4$
\begin{equation}
\begin{aligned}
  \begin{ytableau} {} & {} & {} & {} \end{ytableau}\,,
  \quad \begin{ytableau} {} & {} \\ {} \end{ytableau}\,, 
  \quad \begin{ytableau} {} & {} \end{ytableau}\,, 
  \quad \bullet	
\end{aligned}\label{}
\end{equation}
\item $N$, $\bar{N}=5$
\begin{equation}
\begin{aligned}
\begin{ytableau} {} & {} & {} & {} & {} \end{ytableau}\,,
  \quad \begin{ytableau} {} & {} & {} \\ {} \end{ytableau}\,,
  \quad \begin{ytableau} {} & {} & {} \end{ytableau}	\,,
  \quad \begin{ytableau} {} & {} \\ {}\end{ytableau}\,,
  \quad \begin{ytableau} {} \\ {}\end{ytableau}\,,
  \quad \begin{ytableau} {} \end{ytableau}
\end{aligned}\label{}
\end{equation}
\item $N$, $\bar{N}=6$
\begin{equation}
\begin{aligned}
  &\begin{ytableau} {} & {} & {} & {} & {} & {} \end{ytableau}\,,
  \quad \begin{ytableau} {} & {} & {} & {} \\  {} \end{ytableau}\,,
  \quad \begin{ytableau} {} & {} & {} & {} \end{ytableau}\,,
  \quad \begin{ytableau} {} & {} & {} \\ {} \end{ytableau}\,,
  \quad \begin{ytableau} {} & {} \\ {} & {} \end{ytableau}\,,
  \\
  &\quad \begin{ytableau} {} & {} & {} \end{ytableau}\,,
  \quad \begin{ytableau} {} & {} \\ {} \end{ytableau}\,,
  \quad \begin{ytableau} {} \\ {} \\ {} \end{ytableau}\,,
  \quad 2~\begin{ytableau} {} & {} \end{ytableau}\,,
  \quad \begin{ytableau} {} \end{ytableau} \,,
  \quad \bullet
\end{aligned}\label{}
\end{equation}
\end{itemize}
These will be extensively used  in section \ref{partition_func} to count the physical degrees of freedom.

\section{Enhanced gauge symmetry for the asymmetric vacuum}\label{sec4}
One of the important stringy features in circle compactification of closed string theory is the gauge symmetry enhancement at the self dual radius. In this section, we will introduce vertex operators for the massless states and examine the enhanced gauge symmetry by analyzing the current algebra for the asymmetric vacuum choice.\footnote{In Appendix \ref{App_A}, we briefly summarize the circle compactification of the usual string theory.}

\subsection{Vertex operators for massless states}
In usual string theory, massless states with nonzero KK momentum or winding number at the self dual radius arise only for $N=0,1$ and $\bar{N}=0,1$. In twisted string theory, in contrast with (\ref{maslessordinarycase}),  the massless states fulfill the relations (\ref{massless}), which at the self dual point read 
\begin{equation}
\begin{aligned}
  k_L & = \pm \frac{2}{\sqrt{\alpha^{\prime}}}(\bar{N}-1)^{\frac{1}{2}}\,,
  \\ 
k_R & = \pm \frac{2}{\sqrt{\alpha^{\prime}}}(1-N)^{\frac{1}{2}}\,.
\end{aligned}\label{}
\end{equation}
The right sector is the same as in (\ref{maslessordinarycase}), $N=0$ or $1$, but $\bar{N}$ is now unbounded above. This implies that there are infinitely many massless higher spin states coming from the left sector excitations. Furthermore, at the self dual radius $r=1$, $k_{L}$ and $k_{R}$ should be proportional to  integers
\begin{equation}
\begin{aligned}
  k_{L}&= \Big(\frac{n}{R} +  \frac{ w R}{\alpha'}\Big) \stackrel{\scriptscriptstyle R\to\sqrt{\alpha'}}{=} \frac{1}{\sqrt{\alpha^{\prime}}} (n+w) = \begin{cases}
  	\frac{1}{\sqrt{\alpha'}} \big(2 w \pm 2\big) & \mathrm{if}~ N=0
  	\\ \frac{1}{\sqrt{\alpha'}} 2w & \mathrm{if}~ N=1 
  \end{cases}
\,,
  \\
  k_{R}&= \Big(\frac{n}{R} -  \frac{ w R}{\alpha'}\Big) \stackrel{\scriptscriptstyle R\to\sqrt{\alpha'}}{=} \frac{1}{\sqrt{\alpha^{\prime}}} (n-w)= \begin{cases}
  	\pm\frac{2}{\sqrt{\alpha'}}  &\,\qquad\quad \mathrm{if}~ N=0
  	\\ 0 &\,\qquad\quad \mathrm{if}~ N=1
  \end{cases}\,.
\end{aligned}\label{V1barN}
\end{equation}

From the operator-state mapping, one can identify the physical states as
\begin{equation}
\begin{aligned}
  |0,0,k^{\mu},k_{L},k_{R}\rangle \cong \ :e^{i(k_{\mu} X^{\mu} +k_{L} X_{L}+k_{R}X_{R})}:\,,
  \\
  \alpha^{A}_{-m} \to i \Big(\frac{2}{\alpha'}\Big)^{\frac{1}{2}} \frac{1}{(m-1)!} \partial^{m} X^{A}\,, \qquad m\geq1\,,
  \\
  \bar{\alpha}^{A}_{m} \to i \Big(\frac{2}{\alpha'}\Big)^{\frac{1}{2}} \frac{1}{(m-1)!} \bar{\partial}^{m} X^{A}\,, \qquad m\geq1\,.
\end{aligned}\label{}
\end{equation}
Here $X^{A}$ is decomposed into the external and compact directions as follows:
\begin{equation}
  X_{R/L}^{A} = \{X_{R/L}^{\mu},X_{R/L}\}\,.
\label{}\end{equation}
Then the vertex operators for a generic massless states at the self dual radius in \eqref{genMasslessStates} are given by
\begin{equation}
\begin{aligned}
  &\text{For}~ N=0\,,
\\
  & \quad \bar{a}_{B_{1}B_{2}\cdots B_{n}} :\big(\bar{\partial} X^{B_{1}}_{L}\big)^{\bar{N}_{1}} \big(\bar{\partial}^{2} X^{B_{2}}_{L}\big)^{\bar{N}_{2}} \cdots \big(\bar{\partial}^{n} X^{B_{n}}_{L}\big)^{\bar{N}_{n}} e^{i (k_{\mu} X^{\mu}+k_{L} X_{L}+k_{R}X_{R})} : \,,
\\
&\text{for}~N=1\,,
\\
  &\quad \epsilon_{A,B_{1}B_{2}\dots B_{n}} :\big(\partial X^{A}_{R}\big)\big(\bar{\partial} X^{B_{1}}_{L}\big)^{\bar{N}_{1}} \big(\bar{\partial}^{2} X^{B_{2}}_{L}\big)^{\bar{N}_{2}} \cdots \big(\bar{\partial}^{n} X^{B_{n}}_{L}\big)^{\bar{N}_{n}} e^{i (k_{\mu} X^{\mu}+k_{L} X_{L}+k_{R}X_{R})} :\,.	
\end{aligned}\label{vertOP}
\end{equation}

Now let us consider the internal part of the vertex operators to study the current algebra. If we ignore the external contributions in \eqref{vertOP}, the generic form of the internal part of the vertex operators for massless states or currents are   
\begin{equation}
\begin{aligned}
  V_{\{0,(\bar{N}_{1},\bar{N}_{2}\dots \bar{N}_{n})\}}	&\sim \ :\big(\bar{\partial} X_{L}\big)^{\bar{N}_{1}} \big(\bar{\partial}^{2} X_{L}\big)^{\bar{N}_{2}} \cdots \big(\bar{\partial}^{n} X_{L}\big)^{\bar{N}_{n}} e^{i (k_{L} X_{L}+k_{R}X_{R})} :\,,
  \\
  V_{\{1,(\bar{N}_{1},\bar{N}_{2}\dots \bar{N}_{n})\}}	& \sim \ :\big({\partial} X_{R}) \big(\bar{\partial} X_{L}\big)^{\bar{N}_{1}} \big(\bar{\partial}^{2} X_{L}\big)^{\bar{N}_{2}} \cdots \big(\bar{\partial}^{n} X_{L}\big)^{\bar{N}_{n}} e^{i (k_{L} X_{L}+k_{R}X_{R})} : \,,
\end{aligned}\label{V0barN}
\end{equation}
where $\bar{N} = \sum_{n} n\bar{N}_{n}$. The currents up to $\bar{N}=2$ are listed in Table \ref{tablemassless_asym}.
The currents in (\ref{V0barN}) are factorized into the left and right sectors    
\begin{equation}
\begin{aligned}
  V_{\{N,(\bar{N}_{1},\bar{N}_{2}\dots \bar{N}_{n})\}}(z,\bar{z}) = V_{R}^{(N)}(z) V_{L}^{(\bar{N}_{1},\bar{N}_{2}\dots \bar{N_{n}})}(\bar{z})\,,  \qquad N=0,1
\end{aligned}\label{}
\end{equation}
where 
\begin{equation}
\begin{aligned}
  V_{R}^{(N)}(z) &=~ (\frac{2i}{\sqrt{\alpha^{\prime}}})^N :\big({\partial} X_{R})^{N} e^{ik_{R}X_{R}} : \ = \begin{cases}
  	:e^{\pm\frac{2i}{\sqrt{\alpha'}} X_{R}}:& N= 0 \\ (\frac{i}{\sqrt{\alpha^{\prime}}})\partial X_{R} & N=1
  \end{cases} \,,
  \\
  V_{L}^{(\bar{N}_{1},\bar{N}_{2}\dots \bar{N_{n}})}(\bar{z}) &\sim ~:\big(\bar{\partial} X_{L}\big)^{\bar{N}_{1}} \big(\bar{\partial}^{2} X_{L}\big)^{\bar{N}_{2}} \cdots \big(\bar{\partial}^{n} X_{L}\big)^{\bar{N}_{n}} e^{i k_{L} X_{L}} :\,,
\end{aligned}\label{new_current}
\end{equation}
and $k_{L}$ and $k_{R}$ satisfy (\ref{V1barN}).

Let us introduce $SU(2)_{L/R}$ currents (\ref{current}) which are  defined as  in usual string theory, 
\begin{equation}
  J^{\pm}_{L/R} =\  : \exp\Big(\pm\frac{2i}{\sqrt{\alpha^{\prime}}}X_{L/R}(z)\Big):\,, \qquad J^{3}_{L/R}= \ :(\frac{i}{\sqrt{\alpha^{\prime}}})\partial X_{L/R}(z):\,.
\label{}\end{equation}
For the twisted string case, the right current is identical to  the usual case.  The left current is much more involved than its counterpart in ordinary string theory. Unfortunately, the full structure of the currents is immensely complicated than the conventional string theory. For simplicity, we want to focus on a particular set of internal current generators defined as
\be
G^{(a,b)}(\bar{z})= \,\,:\big(J^3_L(\bar{z})\big)^a\exp\Big(i \frac{2b}{\sqrt{\alpha^{\prime}}}X_L(\bar{z})\Big):\ .
\label{InternalCurrent}\ee
These  correspond to the states constructed only with $\bar{\alpha}^{25}_{1}$  
\begin{equation}
\begin{aligned}
  \big(\bar{\alpha}^{25}_{1}\big)^{a} |{0,k_L}\rangle_L\,.
\end{aligned}\label{lcvoper}
\end{equation}
The norm of the right currents is always positive, since there is no difference with the ordinary string theory.  As we discussed in the previous section for the left currents the norm depends on the number of creation operators.  In the case of (\ref{lcvoper}) if $a$ is even (odd), then the norm is negative (positive). 

Now we construct the full vertex operators containing  the internal current generators (\ref{InternalCurrent}). According to the value of $N$ and the number of legs in the internal direction , there are three independent classes 
\begin{equation}
\begin{aligned}
  V^{(a,b,\pm)}_{\{0,n\}} &= \Big(\frac{2i}{\sqrt{\alpha'}}\Big)^{n-a} \bar{a}^{(a,b,\pm)}_{25\, 25\, \cdots 25\,\mu_{1}\mu_{2}\dots \mu_{(n- a)}} \bar{\partial} X^{\mu_{1}} \bar{\partial} X^{\mu_{2}} \cdots \bar{\partial} X^{\mu_{(n - a)}} e^{i k_{\mu} x^{\mu}} G^{(a,b)} J^{\pm}_{R}
  \\
  &:=  \Big(\frac{2i}{\sqrt{\alpha'}}\Big)^{n-a} \bar{a}^{(a,b,\pm)}_{\mu_{1}\mu_{2}\dots \mu_{(n- a)}} \bar{\partial} X^{\mu_{1}} \bar{\partial} X^{\mu_{2}} \cdots \bar{\partial} X^{\mu_{(n - a)}} e^{i k_{\mu} x^{\mu}} G^{(a,b)} J^{\pm}_{R}\,,
\end{aligned}\label{}
\end{equation}
\begin{equation}
\begin{aligned}
   V^{(a,b,1)}_{\{1,n\}} &= 2\Big(\frac{2i}{\sqrt{\alpha'}}\Big)^{n-a} \epsilon^{(a,b,1)}_{25,25\,25\dots 25\,\nu_{1}\nu_{2}\dots \nu_{(n-a)}} \bar{\partial} X^{\nu_{1}} \bar{\partial} X^{\nu_{2}} \cdots \bar{\partial} X^{\nu_{(n-a)}} e^{i k_{\mu} x^{\mu}} G^{(a,b)} J^{3}_{R}
   \\
   &:=2\Big(\frac{2i}{\sqrt{\alpha'}}\Big)^{n-a} \bar{\epsilon}^{(a,b)}_{\nu_{1}\nu_{2}\dots \nu_{(n-a)}} \bar{\partial} X^{\nu_{1}} \bar{\partial} X^{\nu_{2}} \cdots \bar{\partial} X^{\nu_{(n-a)}} e^{i k_{\mu} x^{\mu}} G^{(a,b)} J^{3}_{R} \, ,
\end{aligned}\label{}
\end{equation}
and 
\begin{equation}
\begin{aligned}
   V^{(a,b,0)}_{\{1,n\}} &=  \Big(\frac{2i}{\sqrt{\alpha'}}\Big)^{n-a+1}\epsilon^{(a,b,0)}_{\mu,25\,25\dots 25\,\nu_{1}\nu_{2}\dots \nu_{(n-a)}} \partial X^{\mu} \bar{\partial} X^{\nu_{1}} \bar{\partial} X^{\nu_{2}} \cdots \bar{\partial} X^{\nu_{(n-a)}} e^{i k_{\mu} x^{\mu}} G^{(a,b)} 	
   \\
   &:= \Big(\frac{2i}{\sqrt{\alpha'}}\Big)^{n-a+1}\epsilon^{(a,b)}_{\mu, \nu_{1}\nu_{2}\dots \nu_{(n-a)}} \partial X^{\mu} \bar{\partial} X^{\nu_{1}} \bar{\partial} X^{\nu_{2}} \cdots \bar{\partial} X^{\nu_{(n-a)}} e^{i k_{\mu} x^{\mu}} G^{(a,b)} \,,
\end{aligned}\label{}
\end{equation}
where $\bar{a}_{\mu_{1}\mu_{2}\dots}^{(a,b)}$, $\epsilon_{\mu, \nu_{1}\nu_{2}\dots}^{(a,b)}$ and $\bar{\epsilon}_{\mu_{1}\mu_{2}\dots}^{(a,b)}$ are polarization tensors for each class. We list the first few levels of massless vertex operators in table \ref{tablemassless_asym}. All the $X_{L}$ and $X_{R}$ contributions have been replaced with the internal currents, $G^{(a,b)}$ and $J_{R}$. 

By definition, the polarization tensors satisfy the following symmetry properties:
\begin{equation}
\begin{aligned}
  \bar{a}_{\mu_{1}\mu_{2}\dots \mu_{n}}^{(a,b)} = \bar{a}_{(\mu_{1}\mu_{2}\dots \mu_{n})}^{(a,b)}\,, \qquad \epsilon_{\mu, \nu_{1}\nu_{2}\dots\nu_{n}}^{(a,b)} = \epsilon_{\mu, (\nu_{1}\nu_{2}\dots\nu_{n})}^{(a,b)}\,, \qquad \bar{\epsilon}_{\mu_{1}\mu_{2}\dots\mu_{n}}^{(a,b)} = \bar{\epsilon}_{(\mu_{1}\mu_{2}\dots\mu_{n})}^{(a,b)}\,.
\end{aligned}\label{}
\end{equation}
They can be interpreted as massless higher-spin fields in the effective field theory point of view. There are a lot of unclear points about their physical properties in this context, such as the no-go theorems for higher-spin theories on a flat background and the higher-spin gauge symmetries \cite{Didenko:2014dwa}. 

\begin{table}[tbp]
\centering
\begin{tabular}{lcc} 
\toprule
Level & Vertex Operator & Classification
\\\midrule
\multirow{2}*{$\begin{matrix}N=0 \\ \bar{N}=1\end{matrix}$} \hspace{0.2cm} & $(\frac{i}{\sqrt{\alpha^{\prime}}})\bar{a}^{(0,0,\pm)}_{\mu}\bar{\partial}X^{\mu}  J^{\pm}_R$ & $\text{\ding{87}}$ \\ \cmidrule{2-3} 
& $\bar{a}^{(1,0,\pm)}G^{(1,0)}  J^{\pm}_R$& $\text{\ding{72}}$
\\ \midrule
\multirow{5}*{$\begin{matrix}N=1\\ \bar{N}=1\end{matrix}$} & $(\frac{i}{\sqrt{\alpha^{\prime}}})^2\epsilon^{(0,0)}_{\mu,\nu}\bar{\partial} X^{\mu} \partial X^{\nu}$ &  \\\cmidrule{2-3} 
& $(\frac{i}{\sqrt{\alpha^{\prime}}})\bar{\epsilon}^{(0,0)}_{\mu}\bar{\partial} X^{\mu} J^3_R$ & $\text{\ding{87}}$ \\\cmidrule{2-3}
& $(\frac{i}{\sqrt{\alpha^{\prime}}})\epsilon^{(1,0)}_{\mu}\partial X^{\mu} G^{(1,0)}$ & \\\cmidrule{2-3}
& $\bar{\epsilon}^{(1,0)}G^{(1,0)} J^3_R $ & $\text{\ding{72}}$ 
\\\midrule
\multirow{4}*{$\begin{matrix}N=0\\ \bar{N}=2\end{matrix}$} & $(\frac{i}{\sqrt{\alpha^{\prime}}})^2 \bar{a}^{(0,\pm1,\pm)}_{\mu\nu}\bar{\partial} X^{\mu} \bar{\partial} X^{\nu}  G^{(0,\pm1)} J^{\pm}_R$ & $\text{\ding{168}}$ \\\cmidrule{2-3} 
& $(\frac{i}{\sqrt{\alpha^{\prime}}}) \bar{a}^{(1,\pm1, \pm)}_{\mu}\bar{\partial} X^{\mu}   G^{(1,\pm1)}J^{\pm}_R$& $\text{\ding{171}}$ \\\cmidrule{2-3}
& $\bar{a}^{(2,\pm1,\pm)}  G^{(2,\pm1)}J^{\pm}_R$ & $\text{\ding{63}}$
\\\midrule
\multirow{8}*{$\begin{matrix}N=1\\ \bar{N}=2\end{matrix}$} & $(\frac{i}{\sqrt{\alpha^{\prime}}})^3 \epsilon^{(0,\pm1)}_{\mu, \nu\rho}  \partial X^{\mu}  \bar{\partial} X^{\nu} \bar{\partial} X^{\rho} G^{(0,\pm1)}$ &  \\\cmidrule{2-3} 
& $(\frac{i}{\sqrt{\alpha^{\prime}}})^2 \epsilon^{(1,\pm1,0)}_{\mu,\nu}\partial X^{\mu} \bar{\partial} X^{\nu}  G^{(1,\pm1)}$ \\\cmidrule{2-3}
& $(\frac{i}{\sqrt{\alpha^{\prime}}}) \epsilon^{(2,\pm1,0)}_{\mu}\partial X^{\mu}   G^{(2,\pm1)} $ &  \\\cmidrule{2-3} 
& $(\frac{i}{\sqrt{\alpha^{\prime}}})^2 \bar{\epsilon}^{(0,\pm1)}_{\mu\nu}\bar{\partial} X^{\mu} \bar{\partial} X^{\nu}  G^{(0,\pm1)} J^3_R$ &  \ding{168}\\\cmidrule{2-3}
& $(\frac{i}{\sqrt{\alpha^{\prime}}}) \bar{\epsilon}^{(1,\pm1)}_{\mu} \bar{\partial} X^{\mu}  G^{(1,\pm1)}J^3_R$ & \ding{171}\\\cmidrule{2-3}
& $\bar{\epsilon}^{(2,\pm1)}  G^{(2,\pm1)}J^{3}_R$ & \ding{63}
\\\midrule
 $\vdots$ & $\vdots$  & $\vdots$ 
\\\bottomrule
\end{tabular}\caption{$25$-dimensional massless spectrum, asymmetric vacuum at the self dual point $R=\sqrt{\alpha^{\prime}}$.}
\label{tablemassless_asym}\end{table}

\subsection{OPE and current algebra}
Before discussing the current algebra and gauge symmetry enhancement, let us first review  the OPE calculation. For computing the OPE, the only necessary information is the two-point correlator. For twisted string theory, the right and left correlators are given by \cite{Lee:2017utr, Huang:2016bdd}
\begin{equation}
\begin{aligned}
  D_R(z_{12}) & = -\frac{\alpha^{\prime}}{2}\log(z_{12})\,,
  \\
  D_L(\bar{z}_{12}) & = +\frac{\alpha^{\prime}}{2}\log(\bar{z}_{12})\,,
\end{aligned}\label{green_asymm}
\end{equation}
where $z_{12} = z_{1}-z_{2}$ and $\bar{z}_{12}=\bar{z}_{1}-\bar{z}_{2}$. Using these correlators, the left and right OPEs  are given by 
\begin{equation}
\begin{aligned}
  X_{R}^{M}(z_{1}) X_{R}^{N}(z_{2})	&= \eta^{MN}D_{R}(z_{12}) + :X_{R}^{N}X_{R}^{M}(z_{2}): + \sum_{k=1}^{\infty} \frac{1}{k!} \big(z_{12}\big)^{k}:X_{R}^{N} X_{R}^{M}(z_{2}): \ ,
  \\
  X_{L}^{M}(\bar{z}_{1}) X_{L}^{N}(\bar{z}_{2})	&=  \eta^{MN}D_{L}(\bar{z}_{12}) + :X_{L}^{N}X_{L}^{M}(\bar{z}_{2}): + \sum_{k=1}^{\infty} \frac{1}{k!} (\bar{z}_{12})^{k}:X_{L}^{N} X_{L}^{M}(\bar{z}_{2}): \ .
\end{aligned}\label{}
\end{equation}

Then, for arbitrary operators $\cal F$ and $\cal G$, the general formula of the OPE for the right sector is 
\begin{equation}
  :{\cal F}_{R} : \ :{\cal G}_{R}:\ = \exp\Big(\int \dd z_{1} \dd z_{2} \ D_{R}(z_{12}) \ \frac{\delta_{\cal F}}{\delta_{\cal F} X^{M}_{R}(z_{1})} \frac{\delta_{\cal G}}{\delta_{\cal G} X_{R M}(z_{2})}\Big) :{\cal F}_{R} \ {\cal G}_{R}: \ ,
\label{}\end{equation}
and for the left sector is 
\begin{equation}
  :{\cal F}_{L} : \ :{\cal G}_{L}:\ = \exp\Big(\int \dd \bar{z}_{1} \dd \bar{z}_{2} \ D_{L}(\bar{z}_{12}) \ \frac{\delta_{\cal F}}{\delta_{\cal F} X^{M}_{L}(\bar{z}_{1})} \frac{\delta_{\cal G}}{\delta_{\cal G} X_{L M}(\bar{z}_{2})}\Big) :{\cal F}_{L}  \ {\cal G}_{L} :\ .
\label{OPE_L}\end{equation}
Since the right correlator is the same as the usual one, the OPE for the right sector coincides with the one presented in Appendix \ref{App_A}. However, the left OPE gets significantly modified due to the sign difference in the left correlator. For instance, the OPE between two left exponential operators, $:\exp(i k_{L} X_L(\bar{z}_1):$ and $:\exp(i k'_{L}X_L(\bar{z}_2)):$\,, can be computed using (\ref{OPE_L}) (\textit{cf}. equation (\ref{OPE1}))
\begin{equation}
\begin{aligned}
:\exp(i k_{L} X_L(&\bar{z}_1): \ : \exp(i k'_{L}X_L(\bar{z}_2)): \ \sim   
\\ 
& \sum_{p=0}^{p<2mn}:\frac{1}{p!}\partial_{\bar{z}_2}^{(p)}\Big(\exp\big(i k_{L}X_L(\bar{z}_2)\big)\Big)\exp\big(i k'_{L}X_L(\bar{z}_2)\big):\big(\bar{z}_{12}\big)^{p-2mn}\,.
\end{aligned}\label{OPE2}
\end{equation}

Now let us move on to the gauge symmetry enhancement by considering  the current algebra. The right currents and their OPEs are exactly the same as in the ordinary case, see Appendix \ref{App_A}. In order to exhibit the adjoint representation of $SU(2)_R$ explicitly, $\{J^\pm_{L/R}, J^3_{L/R}\}$ we have made the pairing using some symbols in table \ref{tablemassless_asym}. For a given $\bar{N}$, each pairing forms a multiplet with respect to the noncompact Lorentz group $SO(1,24)$ and $SU(2)_{R}$.  For example the \ding{87}-pairing corresponds to the vector multiplet with $\bar{N}=1$
\bea
  {}&(\frac{i}{\sqrt{\alpha^{\prime}}})\bar{a}^{(0,0,\pm)}_{\mu}\bar{\partial}X^{\mu}  J^{\pm}_R\, ,\\ \nonumber
 {}&(\frac{i}{\sqrt{\alpha^{\prime}}})\bar{\epsilon}^{(0,0)}_{\mu}\bar{\partial} X^{\mu} J^3_R\, ,
\label{}
\eea
which transform in the adjoint representation of the $SU(2)_R$ group as in the usual circle compactification. As another example, we can pick up the \ding{171}-pairing which forms a vector multiplet with $\bar{N}=2$
\bea\label{gaugefields}
(\frac{i}{\sqrt{\alpha^{\prime}}}) \bar{a}^{(1,\pm1, \pm1)}_{\mu}\bar{\partial} X^{\mu}   G^{(1,\pm1)}J^{\pm}_R\\ \nonumber
(\frac{i}{\sqrt{\alpha^{\prime}}}) \bar{a}^{(1,\pm1,3)}_{\mu} \bar{\partial} X^{\mu}  G^{(1,\pm1)}J^3_R\ .
\eea
While the latter pairing also forms an $SU(2)_R$ triplet, $J^{\pm}_R$ and $J^3_R$, in addition, these vectors contain the left currents, $ G^{(1,\pm1)}$.  

Next we develop the OPE for the left currents. To this end  we compute  the OPE between $G^{(1,0)}$ and $G^{(1,\pm1)}$ defined in (\ref{InternalCurrent})
\begin{equation}
\begin{aligned}
  G^{(1,0)}(\bar{z})  & = \frac{i}{\sqrt{\alpha^{\prime}}}:\partial_{\bar{z}}X_L(\bar{z}): \,,
\\ 
G^{(1,\pm1)}(\bar{z}) & = \frac{i}{\sqrt{\alpha^{\prime}}} :\partial_{\bar{z}} X_{L} \exp\Big(\pm\frac{2i}{\sqrt{\alpha'}}X_L(\bar{z})\Big):
\\
&= \pm \frac{1}{2}\partial_{\bar{z}}\big[:\exp\Big(\pm\frac{2i}{\sqrt{\alpha^{\prime}}}X_L(\bar{z})\Big):\big]\ . 	
\end{aligned}\label{}
\end{equation}
Note that these currents can be written in the exact form, $\partial_{\bar{z}}(\cdots)$. Then the OPEs between these generators are
\begin{equation}
\begin{aligned}
G^{(1,0)}(\bar{z}_1)G^{(1,0)}(\bar{z}_2) & =  -\frac{1}{2\bar{z}_{12}^2}\,,
\\
G^{(1,0)}(\bar{z}_1)G^{(1,\pm1)}(\bar{z}_2) & =  -\frac{1}{2\bar{z}_{12}^{2}} G^{(0,\pm1)}(\bar{z}_2)\mp \frac{1}{\bar{z}_{12}} G^{(1,\pm1)}(\bar{z}_2)\,,
\\
G^{(1,\pm1)}(\bar{z}_1)G^{(1,\pm1)}(\bar{z}_2) & \sim  -\frac{1}{2}\partial_{\bar{z}_1}\partial_{\bar{z}_2}\big[- \frac{1}{2\bar{z}_{12}^2}G^{(0,\pm2)}(\bar{z}_2) \mp \frac{1}{\bar{z}_{12}} G^{(1,\pm2)}(\bar{z}_2) \big] \,,
\\  
 G^{(1,+1)}(\bar{z}_1)G^{(1,-1)}(\bar{z}_2) & \sim \text{regular terms}\,,
\end{aligned}\label{alge_ope} 
\end{equation}
where we have used (\ref{OPE}), (\ref{OPE2}) and the left Green function (\ref{green_asymm}). Obviously, this algebra is not closed over the massless states as in ordinary string theory. Even though we started from $G^{(1,0)}$ and $G^{(1,\pm 1)}$ only, the new currents $G^{(0,1)}$ and $G^{(0,\pm2)}$ arise on the right-hand side of (\ref{alge_ope}). These new currents appear in some massive states. Thus the OPE between massless states may generate massive states, and massive and massless states get mixed. From this fact we can conclude that the enhancement of the gauge symmetry occurs in a  completely different manner compared to usual string theory, where there is no mixing among the massless states and the massive ones. If there were enhancement of the gauge symmetry in twisted string theory,  massive and massless generators would be involve in it. At this level  it is not clear what is the left  current algebra. To figure out the full structure of the algebra, all the currents should be considered . 

Nevertheless, we can find a closed gauge algebra for the zero mode sector of \eqref{alge_ope} (QFT approximation).  After expanding the currents as in  the appendix \ref{App_A}, 
\be
G^{(a,b)}(\bar{z})=\sum_{m=-\infty}^{\infty}\frac{g_m^{(a,b)}}{\bar{z}^{m+1}}\quad \Rightarrow \quad g^{(a,b)}_m=\frac{1}{2\pi i}\int_{\gamma}d\bar{z}\bar{z}^mG^{(a,b)}(\bar{z})\ ,
\ee
from (\ref{alge_ope}) we get the infinite dimensional current algebra 
\bea
\big[g_m^{(1,0)},g_n^{(1,0)}\big] & = & -\frac{m}{2} \delta_{0,m+n}\\ \nonumber
\big[g_m^{(1,0)},g_n^{(1,\pm1)}\big] & = & -\frac{1}{2}m\ g_m^{(0,\pm1)}\mp  g^{(1,\pm1)}_{m+n}\\ \nonumber
\big[g_m^{(1,+1)},g_n^{(1,+1)}\big] & = & \frac{mn}{4}\big(g_{m+n-3}^{(0,+2)} + 2ig^{(1,+1)}_{m+n-2}\big)\\ \nonumber
\big[g_m^{(1,-1)},g_n^{(1,-1)}\big] & = & \frac{mn}{4}\big(g_{m+n-3}^{(0,-2)} -2ig^{(1,-1)}_{m+n-2}\big)\\ \nonumber
\big[g_m^{(1,+1)},g_n^{(1,-1)}\big] & = & 0\ .
\eea
For the zero mode component of this algebra we have
\bea
\big[g_0^{(1,0)},g_0^{(1,\pm1)}\big] & = & \mp  g^{(1,\pm1)}_{0}\\ \nonumber
\big[g_0^{(1,+1)},g_0^{(1,-1)}\big] & = & 0\ .
\eea
This is a well known Lie algebra, called the group contraction algebra  \cite{Gilmore:2008zz}. In this case, it is nothing but the contraction of the  $so(3)$ algebra. 

If we consider fields up to $\bar{N}=2$ with $N=0\ , 1$ \footnote{Since at the self dual point there are infinite many massless fields it is not clear what is the mechanism that allows this truncation.} we can identify the enhanced gauge group as $ISO(2)_L \times SU(2)_R$. The corresponding algebra is
\be
su(2)_L \times su(2)_R\sim so(3)_L \times su(2)_R
\xrightarrow[{}]{\text{contract}} iso(2)_L \times su(2)_R\ .
\ee
Note that the gauge symmetry of the effective field theory is determined by the zero mode sector of the current algebra. Hence, even though the full current algebra is not clear, the gauge symmetry of the effective field theory,  which is associated with the zero mode generators only, contains the $ISO(2)_L \times SU(2)_R$ subgroup.

There are several open questions regarding this gauge symmetry enhancement  realization. For instance the vector fields (\ref{gaugefields}) have the unusual structure $\bar{a}_{\mu}^{ab}\tau^a_L\tau^b_R$, being $\tau^a_L$ and $\tau^b_R$ the generators in each group. Regarding these fields as gauge fields is unclear and needs further exploration.


\section{The partition function}\label{partition_func}
It is interesting to see how the above results in the canonical approach can be applied to the partition function analysis. The  partition function for this string theory on a flat non compact space was already presented in \cite{Lee:2017utr}. Before continuing  we want to make some comments about the absence of modular invariance \footnote{In \cite{Casali:2017mss, Casali:2017zkz} was suggested that a complexification of the moduli $(\tau_1, \tau_2)\rightarrow (\xi_1,\xi_2)\in \mathbb{C}^2$  complemented with an appropriated  integration cycle choice could restore the modular invariance.  In this case the physical interpretation  is obscure but $Z(\xi_1,\xi_2)$ certainly  would be convergent.} For a review of the calculation of the partition function see Appendix \ref{appB}. 
 
 In usual bosonic  string theory the  modular invariance  is achieved, because  the left and the right contribution to the partition function $Z(\tau,\bar{\tau})$ can be written in terms of the Dedekind function. The Dedekind function transforms properly under $\tau\rightarrow \tau+1$ and $\tau\rightarrow -\frac{1}{\tau}$ in such a way that
 \be
 Z=\int \frac{d\tau d\bar{\tau}}{\tau_2}Z(\tau,\bar{\tau})
 \ee
 is invariant. Due to this invariance the integration over the moduli space is performed on the fundamental domain. 
The Dedekind function, on the one hand, is defined over the upper half plane $\tau = \tau_1 +i \tau_2$ for $\tau_2 > 0$ by
\be
\eta(\tau)= q^{\frac{1}{24}} \prod_{k=1}^{\infty}(1-q^k),
\ee
where $q=e^{2 \pi i \tau}$.
Alternatively, one can use the Euler generating function to write the Dedekind function as 
\be
\eta(\tau)^{-1}= q^{-\frac{1}{24}}\frac{1}{ \prod_{k=1}^{\infty}(1-q^k)}=q^{-\frac{1}{24}}\sum_{n=0}^{\infty}p(n)q^n, \label{dede}
\ee
where $p(n)$ are the number of partitions for each level.
This chain of equalities in (\ref{dede}) exists, because the infinite product and series converge in the upper half plane.
 
On the other hand, in the calculation of $Z(\tau,\bar{\tau})$ for the asymmetric vacuum one finds that the right moving contribution is the usual one as in ordinary string theory. However, the left moving contribution is proportional to
\be
\sum_{n=0}^{\infty}p(n)e^{2\pi i n \bar{\tau}}. \label{nodede}
\ee 
The expression  (\ref{nodede}) looks similar to  the Dedekind function (\ref{dede}), however we should be careful with this identification. Noticed that (\ref{nodede}) is defined over the lower half plane  $\bar{\tau} = \tau_1 -i \tau_2$ for $\tau_2 > 0$.  As a consequence  (\ref{nodede}) should be regarded as a formal expression since the series over the lower half plane does not converge. For obtaining the modular properties of the Dedekind function (\ref{dede}) the convergence of the series is an indispensable requirement. It is also well known that the Dedekind function does not admit any analytical extension to the lower half plane. 

Now let us consider the partition function for the $S^1$ compactification. The partition function takes a form similar to (\ref{parfunc}). The only difference is in the zero mode components of the momenta, i.e. 
\bea\label{parfunc1}
\text{Z}(\tau) & = & \text{e}^{-4\pi\text{i}\tau_1}\\\nonumber
{} & {} &\text{Tr}_{R_0}\big[\text{e}^{ +\pi \text{i}\tau( \alpha_0^2+\frac{1}{2}\hat{P}_{R}^2)}\big]\text{Tr}_R\big[\text{e}^{+2\pi\text{i}\tau\hat{N}_B}\big]\text{Tr}_R\big[(-1)^{F}c_0b_0\text{e}^{+2\pi\text{i}\tau\hat{N}_g}\big]\\ \nonumber
{} & {} &\text{Tr}_{L_0}\big[\text{e}^{-\pi \text{i}\tau(\bar{\alpha}_0^2+\frac{1}{2}\hat{P}_{L}^2)}\big] \text{Tr}_L\big[\text{e}^{-2\pi\text{i}\overline{\tau}\hat{\overline{N}}_B}\big]\text{Tr}_L\big[(-1)^{\overline{F}}\overline{c}_0\overline{b}_0\text{e}^{-2\pi\text{i}\overline{\tau}\hat{\overline{N}}_g}\big]\ . \nonumber
\eea
After tracing over each Hilbert space we get
\begin{equation}
\begin{aligned}
  \text{Z}(\tau) & = \int\frac{\dd^{25}k}{(2\pi)^{25}}e^{-\pi\tau_2 \alpha^{\prime}k^2}\sum_{n,\omega}\Big[ \text{e}^{-\pi\tau_2(\frac{\alpha^{\prime}}{R^2}n^2+\frac{R^2}{\alpha^{\prime}}\omega^2)}\\
{}&\qquad \times\mathrm{e}^{-2\pi\text{i}(n\omega+2)\tau_1} \Big[\sum_{N=0}^{\infty}P\big(N\big)\Big(\text{e}^{2\pi\text{i}\tau}\Big)^{N}\Big]^{24} \Big[\sum_{\overline{N}=0}^{\infty}P\big(\overline{N}\big)\Big(\text{e}^{2\pi\text{i}\overline{\tau}}\Big)^{\overline{N}}\Big]^{24}\ .	
\end{aligned}\label{Z_Hilbert}
\end{equation}
Note that the integration along the circle direction, $k^{25}$, is replaced by the summation over the momentum $n$ and winding number $w$. 

As we showed in section \ref{sec:bosonic}, infinitely many states arise due to the winding modes. Interestingly enough, the one-loop partition function encodes this information, and one can single out the contribution of a given $n$ and $w$ from the partition function 
\begin{equation}
\begin{aligned}
  \mathrm{Z}_{n,w}(\tau) &= \int\frac{\dd^{25}k}{(2\pi)^{25}}e^{-\pi\tau_2 \alpha^{\prime}k^2}\text{e}^{-\pi\tau_2\big(\frac{\alpha^{\prime}}{R^2}n^2+\frac{R^2}{\alpha^{\prime}}\omega^2\big)} \text{e}^{-2\pi\text{i}(n\omega+2)\tau_1} \\
{}&\qquad\times\Big[\sum_{N=0}^{n\omega+2}P\big(N\big)\Big(\text{e}^{2\pi\text{i}\tau}\Big)^{N}\Big]^{24}\Big[\sum_{\overline{N}=0}^{n\omega+2}P\big(\overline{N}\big)\Big(\text{e}^{2\pi\text{i}\overline{\tau}}\Big)^{\overline{N}}\Big]^{24}\,.
\end{aligned}\label{Znwt}
\end{equation}
Then the total one loop partition function can be recast as a sum over all the possible momentum and winding numbers
\bea
\text{Z}(\tau) & \sim & \sum_{n,w \atop n\omega+2\geq 0} \mathrm{Z}_{n,w}(\tau) \ .
\eea
As explained before, this partition function is not modular invariant \cite{Lee:2017utr}. Thus, one can perform the integration along the $\tau_1$ direction explicitly,
\begin{equation}
  \mathrm{Z}_{n,w}(\tau_{2}) = \int_{-\frac{1}{2}}^{\frac{1}{2}} \dd \tau_{1}\, \text{Z}_{n,w}(\tau)\,,
\label{Znwt2}\end{equation}
and compare with the corresponding field theory partition function.

In the following subsections, we will explicitly examine the correspondence between the field theory and string theory partition functions for the first few levels.

\subsection{$n=0$ or $w=0$}
In this case, the level matching constraint is equivalent to the non-compact spacetime case (\ref{LVmatching}),
\begin{equation}
  N+\bar{N} = 2\,.
\label{}\end{equation}
Since the results between $n=0$ and $w=0$ are almost identical, we focus only on the $n=0$ case. There are three types  of states that satisfy  the level matching constraint:
\begin{equation}
\begin{aligned}
  1.~~ &N=0\,, \quad \bar{N}=2\,, \quad M^{2} = \frac{1}{\alpha'} \frac{n^{2}}{r^{2}}-\frac{4}{\alpha'}\,, \quad &\Big|\psi^{N+\bar{N}=2}_{(0,2)} \Big\rangle \,,
  \\
  2.~~&N=1\,, \quad \bar{N}=1\,, \quad M^{2} = \frac{1}{\alpha'}\frac{n^{2}}{r^{2}}\,, \quad &\Big|\psi^{N+\bar{N}=2}_{(1,1)}\Big\rangle \,,
  \\
  3.~~&N=2\,, \quad \bar{N}=0\,, \quad M^{2} = \frac{1}{\alpha'}\frac{n^{2}}{r^{2}}+\frac{4}{\alpha'}\,, \quad &\Big|\psi^{N+\bar{N}=2}_{(2,0)}\Big\rangle \,.
\end{aligned}\label{}
\end{equation}
Note that the first state becomes massless  for $n=2$ at the self-dual radius, $r=1$.

The first and third states are represented by the traceless symmetric rank 2 tensor of the $SO(D-1)$ little group. It is encoded by the Young tableau
\begin{equation}
\begin{aligned}
  \Big|\psi^{N+\bar{N}=2}_{(0,2)} \Big\rangle ~:~ \bullet \otimes ~\begin{ytableau} {} & {} \end{ytableau} \,, 
  \\ 
  \Big|\psi^{N+\bar{N}=2}_{(2,0)}\Big\rangle ~:~ \begin{ytableau} {} & {} \end{ytableau} ~\otimes \bullet\,.
\end{aligned}\label{}
\end{equation}
The number of degrees of freedom for each state is given by the dimension of the Young tableau,
\begin{equation}
  \dim \Big[~\begin{ytableau} {} & {} \end{ytableau}~\Big] = \frac{(D+1)(D-2)}{2}.
\label{dimL2_2_0}\end{equation}
The second state is represented by 
\begin{equation}
  \begin{ytableau} *(gray){} \end{ytableau} \otimes \begin{ytableau} *(gray){} \end{ytableau}
\label{grayb}\end{equation}
where the grey box denotes the $SO(D-2)$ representation of the little group which is relevant to the massless states. As in the open string, the level 1 state, $N=1$ or $\bar{N}=1$, corresponds to the massless vector boson. Therefore, a single box state always comes with a  $SO(D-2)$ representation.
The dimension of the state (\ref{grayb}) is given by
\begin{equation}
  \dim\Big[~\begin{ytableau} *(gray){} \end{ytableau} \otimes \begin{ytableau} *(gray){} \end{ytableau}~\Big] = (D-2) \times (D-2)\,.
\label{dimL2_1_1}\end{equation}
Combining (\ref{dimL2_2_0}) and (\ref{dimL2_1_1}) and using the QFT definition of the partition function, 
\begin{equation}
Z(s)=\int \frac{\dd^{D-1} k}{(2\pi)^{(D-1)}}e^{- s  k^2}\sum_i d_i e^{-m_i^2  s}
\end{equation}
with degeneracy $d_i$ given by the dimension of the representations, we have
\begin{equation}
\begin{aligned}
  \mathrm{Z}_{n,0}(\tau_{2}) &= \int \frac{\dd^{D-1} k}{(2\pi)^{(D-1)}} e^{-\pi\tau_2 \alpha^{\prime}k^2} \Big[ ~ (D-2)^{2} e^{-\frac{\pi  n^2 \tau_{2}}{r^2}}
  \\
  &\qquad\qquad\qquad +\frac{(D-2)(D+1)}{2} \left( e^{-\frac{\pi  n^2 \tau_{2}}{r^2}-4 \pi  \tau_2} +e^{4 \pi \tau_{2}-\frac{\pi  n^2 \tau_{2}}{r^2}} \right)~\Big]\ ,
\end{aligned}\label{fieldZ11}
\end{equation}
here,  $s=\pi \alpha^{\prime} \tau_2$ is Schwinger parameter. 
On the other hand, using the twisted string  partition function (\ref{Znwt}) with (\ref{Znwt2}) at the critical dimension $D=26$ we get
\begin{equation}
  \mathrm{Z}_{n,0} (\tau_{2})= \int \frac{\dd^{25} k}{(2\pi)^{25}} e^{-\pi\tau_2 \alpha^{\prime}k^2} \Big[ ~ 576 e^{-\frac{\pi  n^2 \tau_{2}}{r^2}}+324 e^{-\frac{\pi  n^2 \tau_{2}}{r^2}-4 \pi  \tau_2}+324 e^{4 \pi \tau_{2}-\frac{\pi  n^2 \tau_{2}}{r^2}}\Big]\ .
\label{zzzz}
\end{equation}

These two partition functions have been computed using different definitions. Nevertheless it is remarkable that they coincide for $D=26$. In what follows we will see this correspondence level by level.

\subsection{$n=\pm1$ and $w=\pm1$}
In this case, the level matching constraint is given by $N+\bar{N}=3$, and there are $4$ states 
\begin{equation}
\begin{aligned}
  1.~~ &N=0\,, \quad \bar{N}=3\,, \quad M^{2} = \frac{1}{\alpha'}\left(r+\frac{1}{r}\right)^{2}-\frac{6}{\alpha'}\,, \quad &\Big|\psi^{N+\bar{N}=3}_{(0,3)} \Big\rangle \,,
  \\
  2.~~&N=1\,, \quad \bar{N}=2\,, \quad M^{2} = \frac{1}{\alpha'}\left(r+\frac{1}{r}\right)^{2}-\frac{4}{\alpha'}\,, \quad &\Big|\psi^{N+\bar{N}=3}_{(1,2)}\Big\rangle \,,
  \\
  3.~~&N=2\,, \quad \bar{N}=1\,, \quad M^{2} = \frac{1}{\alpha'}\left(r+\frac{1}{r}\right)^{2}\,, \quad &\Big|\psi^{N+\bar{N}=3}_{(2,1)}\Big\rangle \,,
  \\
  4.~~&N=3\,, \quad \bar{N}=0\,,	 \quad M^{2} = \frac{1}{\alpha'}\left(r+\frac{1}{r}\right)^{2}+\frac{4}{\alpha'}\,, \quad &\Big|\psi^{N+\bar{N}=3}_{(3,0)}\Big\rangle \,.
\end{aligned}\label{n1w1}
\end{equation}
As we observed in (\ref{NbN3}), at the self-dual radius $r=1$, the second state $\Big|\psi^{N+\bar{N}=3}_{(1,2)}\Big\rangle$ becomes massless. 
  
The first and the last states in (\ref{n1w1}) have the following tensor representations of  the $SO(D-1)$ little group,
\begin{equation}
\Big|\psi^{N+\bar{N}=3}_{(0,3)} \Big\rangle :~ \bullet\otimes
\Big(~\begin{ytableau} {} & {} & {} \end{ytableau}
+
\begin{ytableau} {}\\{} \end{ytableau}~\Big)\,,
\label{}\end{equation}
and 
\begin{equation}
\Big|\psi^{N+\bar{N}=3}_{(3,0)} \Big\rangle :~
\Big(~\begin{ytableau} {} & {} & {} \end{ytableau}
+
\begin{ytableau} {}\\{} \end{ytableau}~\Big)\otimes \bullet\,.
\label{}\end{equation}
The number of states of each irreducible representation is 
\begin{equation}
  \dim\Big[~\begin{ytableau}
 {} & {} & {}
\end{ytableau}~\Big] = \frac{(D+3) (D-2) (D-1)}{6}\,, \qquad \dim\Big[~\begin{ytableau}
 {}\\{}
\end{ytableau}~\Big] = \frac{(D-1) (D-2)}{2}\,.
\label{}\end{equation}
The number of degrees of freedom for the other states is
\begin{equation}
\begin{aligned}
  \mathrm{d.o.f}\Big(~\Big|\psi^{N+\bar{N}=3}_{(0,3)} \Big\rangle~\Big) = \mathrm{d.o.f}\Big(	~\Big|\psi^{N+\bar{N}=3}_{(3,0)} \Big\rangle~\Big) = \frac{(D-2) (D-1) (D+6)}{6} \,.
\end{aligned}\label{}
\end{equation}
The second and third states are represented by 
\begin{equation}
\begin{ytableau}
 *(gray){}
\end{ytableau}
\,\otimes\,
\begin{ytableau}
 {} & {} 
\end{ytableau}\,, \qquad\text{and }\qquad 
\begin{ytableau}
 {} & {}
\end{ytableau}
\,\otimes\,
\begin{ytableau}
 *(gray) {} 
\end{ytableau}\,,
\label{}\end{equation}
respectively. The dimension of these representations with respect to the little group is given by
\begin{equation}
\begin{aligned}
    \dim\Big[~\begin{ytableau}
 *(gray){}
\end{ytableau}
\,\otimes\,
\begin{ytableau}
 {} & {}
\end{ytableau}~\Big] = (D-2) \times \frac{(D-2) (D+1)}{2}\,,
\end{aligned}\label{}
\end{equation}
and the number of degrees of freedom for these states is
\begin{equation}
    \mathrm{d.o.f}\Big(~\Big|\psi^{N+\bar{N}=3}_{(1,2)} \Big\rangle~\Big) = \mathrm{d.o.f}\Big(	~\Big|\psi^{N+\bar{N}=3}_{(2,1)} \Big\rangle~\Big) = \frac{(D-2)^{2} (D+1)}{2}\ .
\label{}\end{equation}

Combining these results, we have the QFT partition function for the $n=w=\pm1$ sector 
\begin{equation}
\begin{aligned}
  \mathrm{Z}_{1,1}(\tau_2) &= \int \frac{\dd^{D-1} k}{(2\pi)^{(D-1)}} e^{-\pi\tau_2 \alpha^{\prime}k^2} 
  \\
  &\qquad \times \Bigg[\frac{(D-2) (D-1)(D+6)}{6} \Big(e^{-\pi \tau_{2} \left(r+\frac{1}{r}\right)^2 +8\pi \tau_{2}} + e^{-\pi \tau_{2}\left(r+\frac{1}{r}\right)^2 -4\pi \tau_{2}} \Big) 
  \\
  &\qquad\qquad + \frac{(D-2)^{2} (D+1)}{2}  \left(e^{-\pi\tau_{2}\left(r+\frac{1}{r}\right)^2} + e^{-\pi\tau_{2}\left(r+\frac{1}{r}\right)^2 +4\pi\tau_{2}} \right)~\Bigg]\,.
\end{aligned}\label{z11}
\end{equation}
On the other hand, the evaluation of  (\ref{Znwt2}) at the critical dimension, $D=26$, gives
\begin{equation}
\begin{aligned}
  Z_{1,1}(\tau_{2}) &= \int_{-\frac{1}{2}}^{\frac{1}{2}} \dd \tau_{1} \mathrm{Z}_{1,1} (\tau) 
  \\
  &= \int \frac{\dd^{25} k}{(2\pi)^{25}} e^{-\pi\tau_2 \alpha^{\prime}k^2} 32 \Big[~100 \Big(e^{-\pi \tau_{2} \left(r+\frac{1}{r}\right)^2 +8\pi \tau_{2}} + e^{-\pi \tau_{2}\left(r+\frac{1}{r}\right)^2 -4\pi \tau_{2}}\Big)  
  \\
  &\qquad\qquad +243 \Big( e^{-\pi\tau_{2}\left(r+\frac{1}{r}\right)^2 } + e^{-\pi\tau_{2}\left(r+\frac{1}{r}\right)^2 +4\pi\tau_{2}}\Big) ~\Big] \,.
\end{aligned}\label{}
\end{equation}
This is exactly the same result as in  the field theory (\ref{z11}) with $D=26$.

\subsection{$n=\pm2$ and $w=\pm2$} 
In this case, the level matching constraint is given by $N+\bar{N} = 6$, and there are $7$  states:
\begin{equation}
\begin{aligned}
  1.~~ &N=0\,, \quad \bar{N}=6\,, \quad M^{2} = \frac{4}{\alpha'}\left(r+\frac{1}{r}\right)^{2}-\frac{20}{\alpha'}\,, \quad &\Big|\psi^{N+\bar{N}=6}_{(0,6)} \Big\rangle \,,
  \\
  2.~~&N=1\,, \quad \bar{N}=5\,, \quad M^{2} = \frac{4}{\alpha'}\left(r+\frac{1}{r}\right)^{2}-\frac{16}{\alpha'}\,,\quad &\Big|\psi^{N+\bar{N}=6}_{(1,5)} \Big\rangle \,,
  \\
  3.~~&N=2\,, \quad \bar{N}=4\,, \quad M^{2} = \frac{4}{\alpha'}\left(r+\frac{1}{r}\right)^{2}-\frac{12}{\alpha'}\,,\quad &\Big|\psi^{N+\bar{N}=6}_{(2,4)} \Big\rangle \,,
  \\
  4.~~&N=3\,, \quad \bar{N}=3\,,	 \quad M^{2} = \frac{4}{\alpha'}\left(r+\frac{1}{r}\right)^{2}-\frac{8}{\alpha'}\,,\quad &\Big|\psi^{N+\bar{N}=6}_{(3,3)} \Big\rangle \,,
  \\
  5.~~&N=4\,, \quad \bar{N}=2\,,	 \quad M^{2} = \frac{4}{\alpha'}\left(r+\frac{1}{r}\right)^{2}-\frac{4}{\alpha'}\,,\quad &\Big|\psi^{N+\bar{N}=6}_{(4,2)} \Big\rangle \,,
  \\
  6.~~&N=5\,, \quad \bar{N}=1\,,	 \quad M^{2} = \frac{4}{\alpha'}\left(r+\frac{1}{r}\right)^{2}\,,\quad &\Big|\psi^{N+\bar{N}=6}_{(5,1)} \Big\rangle \,,
  \\
  7.~~&N=6\,, \quad \bar{N}=0\,,	 \quad M^{2} = \frac{4}{\alpha'}\left(r+\frac{1}{r}\right)^{2}+\frac{4}{\alpha'}\,.\quad &\Big|\psi^{N+\bar{N}=6}_{(6,0)} \Big\rangle \,.
\end{aligned}\label{}
\end{equation}

The first and the last states consist of the following irreducible representations:
\begin{equation}
\begin{aligned}
  \Big|\psi^{N+\bar{N}=6}_{(0,6)} \Big\rangle &:~ \bullet \otimes\Bigg(\begin{ytableau}
  {} & {} & {} & {} & {} & {} \end{ytableau}
  + \begin{ytableau} {} & {} & {} & {} \\  {} \end{ytableau}
  + \begin{ytableau} {} & {} & {} & {} \end{ytableau}
  + \begin{ytableau} {} & {} & {} \\ {} \end{ytableau}
  + \begin{ytableau} {} & {} \\ {} & {} \end{ytableau}
  \\
  &\qquad\qquad +\begin{ytableau} {} & {} & {} \end{ytableau}
  + \begin{ytableau} {} & {} \\ {} \end{ytableau}
  + \begin{ytableau} {} \\ {} \\ {} \end{ytableau}
  + 2~\begin{ytableau} {} & {} \end{ytableau}
  + \begin{ytableau} {} \end{ytableau} 
  + \bullet~\Bigg) \ ,
\end{aligned}\label{}
\end{equation}
and 
\begin{equation}
\begin{aligned}
  \Big|\psi^{N+\bar{N}=6}_{(6,0)} \Big\rangle &:~  \Bigg(\begin{ytableau}
  {} & {} & {} & {} & {} & {} \end{ytableau}
  + \begin{ytableau} {} & {} & {} & {} \\  {} \end{ytableau}
  + \begin{ytableau} {} & {} & {} & {} \end{ytableau}
  + \begin{ytableau} {} & {} & {} \\ {} \end{ytableau}
  + \begin{ytableau} {} & {} \\ {} & {} \end{ytableau}
  \\
  &\qquad\qquad +\begin{ytableau} {} & {} & {} \end{ytableau}
  + \begin{ytableau} {} & {} \\ {} \end{ytableau}
  + \begin{ytableau} {} \\ {} \\ {} \end{ytableau}
  + 2~\begin{ytableau} {} & {} \end{ytableau}
  + \begin{ytableau} {} \end{ytableau} 
  + \bullet~\Bigg)\otimes \bullet \ .
\end{aligned}\label{}
\end{equation}
Using the result in (\ref{dimYD}), we can obtain the number of the physical degrees of freedom,
\begin{equation}
\begin{aligned}
  \mathrm{d.o.f}\Big[~\Big|\psi^{N+\bar{N}=6}_{(0,6)} \Big\rangle~\Big] &=\mathrm{d.o.f}\Big[~\Big|\psi^{N+\bar{N}=6}_{(6,0)} \Big\rangle~\Big]
  \\
  &= \frac{(D-2)(D-1)(D+8)(D^3+28D^2+57D-90)}{6!} 	\,.
\end{aligned}\label{}
\end{equation}

Next, the second and the sixth states are represented as follows:
\begin{equation}
\begin{aligned}
\Big|\psi^{N+\bar{N}=6}_{(1,5)} \Big\rangle &=\begin{ytableau}
 *(gray){}
\end{ytableau}
\,\otimes\,\Big(~
\begin{ytableau}
 {} & {} & {} & {} & {}
\end{ytableau}
+
\begin{ytableau}
 {} & {} & {} \\ {}
\end{ytableau}
+
\begin{ytableau}
 {} & {} & {} 
\end{ytableau}
+
\begin{ytableau}
 {} & {} \\ {}
\end{ytableau}
+
\begin{ytableau}
 {} \\ {}
\end{ytableau}
+
\begin{ytableau}
 {} 
\end{ytableau}~\Big)\,,
\end{aligned}\label{}
\end{equation}
and 
\begin{equation}
\begin{aligned}
\Big|\psi^{N+\bar{N}=6}_{(5,1)} \Big\rangle &=
\Big(~ \begin{ytableau} {} & {} & {} & {} & {} \end{ytableau}
+ \begin{ytableau} {} & {} & {} \\ {} \end{ytableau}
+ \begin{ytableau} {} & {} & {} \end{ytableau}		
+ \begin{ytableau} {} & {} \\ {}\end{ytableau}
+ \begin{ytableau} {} \\ {}\end{ytableau} 		
+ \begin{ytableau} {} \end{ytableau}~\Big)
\,\otimes\,\begin{ytableau} *(gray){} \end{ytableau}\,.																		
\end{aligned}\label{}
\end{equation}
As before, the grey box denotes  the vector representation of the $SO(d-2)$ little group.
From (\ref{dimYD}), we obtain the number of degrees of freedom for these states,
\begin{equation}
\begin{aligned}
    \mathrm{d.o.f}\Big[~\Big|\psi^{N+\bar{N}=6}_{(1,5)} \Big\rangle~\Big] &= \mathrm{d.o.f}\Big[~\Big|\psi^{N+\bar{N}=6}_{(5,1)} \Big\rangle~\Big] 
    \\
    &= \frac{(D-2)^2 (D+1) (D+4) (D (D+17)-30)}{5!} \,.
\end{aligned}\label{}
\end{equation}
The third and fifth states are described by the following irreducible representations, 
\begin{equation}
  \Big|\psi^{N+\bar{N}=6}_{(2,4)} \Big\rangle : \begin{ytableau}
 {} & {}
\end{ytableau}
\,\otimes\,
\Big(~\begin{ytableau}
 {} & {} & {} & {} 
\end{ytableau} + \begin{ytableau}
  {} & {} \\ {}
\end{ytableau} + \begin{ytableau}
 {} & {} 
\end{ytableau} + \bullet ~\Big) \ ,
\label{}\end{equation}
and 
\begin{equation}
\Big|\psi^{N+\bar{N}=6}_{(4,2)} \Big\rangle : \Big(~\begin{ytableau}
 {} & {} & {} & {} 
\end{ytableau} + \begin{ytableau}
  {} & {} \\ {}
\end{ytableau} + \begin{ytableau}
 {} & {} 
\end{ytableau} + \bullet ~\Big)
\otimes
\begin{ytableau}
 {} & {}
\end{ytableau}\ .
\label{}
\end{equation}\
Their number of degrees of freedom is given by
\begin{equation}
  \mathrm{d.o.f}\Big[~\Big|\psi^{N+\bar{N}=6}_{(2,4)} \Big\rangle~\Big] = \mathrm{d.o.f}\Big[~\Big|\psi^{N+\bar{N}=6}_{(4,2)} \Big\rangle~\Big] =\frac{(D-2) (D-1) (D+1) (D+12)}{4!} \,.
\label{}\end{equation}
Finally, we consider the fourth state, $N=\bar{N}=3$. It is represented by
\begin{equation}
\Big|\psi^{N+\bar{N}=6}_{(3,3)} \Big\rangle : 
\Big(~\begin{ytableau} {} & {} & {} \end{ytableau} 
+ \begin{ytableau} {}\\{}\end{ytableau}~\Big) 
\otimes \Big(~\begin{ytableau}
 {} & {} & {}
\end{ytableau} + 
\begin{ytableau}
 {}\\{}
\end{ytableau}~\Big)\,.
\label{}\end{equation}
The number of physical degrees of freedom for this state is given by
\begin{equation}
  \mathrm{d.o.f}\Big[~\Big|\psi^{N+\bar{N}=6}_{(3,3)} \Big\rangle~\Big] = \left(\frac{(D-2)(D-1)(D+6)}{3!} \right)^{2} \ .
\label{}\end{equation}

Collecting all the physical states we can construct the corresponding field theory partition function 
\begin{equation}
\begin{aligned}
  \mathrm{Z}_{2,2}(\tau_{2})&= \int \frac{\dd^{D-1} k}{(2\pi)^{(D-1)}} e^{-\pi\tau_2 \alpha^{\prime}k^2}  
  \\
  &\times \Big[\textstyle\frac{(D-2)(D-1)(D+8)(D^3+28D^2+57D-90)}{6!}\, e^{-4\pi\tau_{2} \left(r+\frac{1}{r}\right)^{2} } \Big(e^{20\pi\tau_2}+e^{-4\pi\tau_2}\Big)
  \\
  &\qquad \textstyle+ \frac{(D-2)^2 (D+1) (D+4) (D (D+17)-30)}{5!}\, e^{-4\pi\tau_{2} \left(r+\frac{1}{r}\right)^{2}} \Big(1+e^{16\pi\tau_2}\Big)
  \\
  &\qquad \textstyle+\frac{(D-2) (D-1) (D+1) (D+12)}{4!}\, e^{-4\pi\tau_{2}\left(r+\frac{1}{r}\right)^{2}}  \Big(e^{8\pi\tau_2}+e^{4\pi\tau_2}\Big)
  \\
  &\qquad \textstyle+\left(\frac{(D-2)(D-1)(D+6)}{3!} \right)^{2}e^{8\pi\tau_2-4\pi\tau_{2}\left(r+\frac{1}{r}\right)^{2}} ~\Big]\,.
\end{aligned}\label{Z22_FT}
\end{equation}
At the critical dimension the explicit integration in $\tau_1$  (\ref{Znwt2}) is given by
\begin{equation}
\begin{aligned}
  Z_{2,2}&(\tau_{2}) = Z_{-2,-2} (\tau_{2})  
    \\
    &=\int \frac{\dd^{25} k}{(2\pi)^{25}} \Big[ e^{-\pi\tau_2 \alpha^{\prime}k^2}  1073720\, e^{-4\pi\tau_2-4\pi\tau_{2} \left(r+\frac{1}{r}\right)^{2} } + 4230144\, e^{-4\pi\tau_{2}\left(r+\frac{1}{r}\right)^{2}}
    \\
    &\quad +8310600\, e^{4\pi\tau_2 -4\pi\tau_{2}\left(r+\frac{1}{r}\right)^{2}}+10240000\,e^{8\pi\tau_2-4\pi\tau_{2}\left(r+\frac{1}{r}\right)^{2}}
    \\
    &\quad +8310600\, e^{12\pi\tau_2-4\pi\tau_{2}\left(r+\frac{1}{r}\right)^{2}} +4230144\, e^{16\pi\tau_2 -4\pi\tau_{2}\left(r+\frac{1}{r}\right)^{2}}
    \\
    &\quad +1073720\, e^{20\pi\tau_2 -4\pi\tau_{2}\left(r+\frac{1}{r}\right)^{2}}\Big]\,,
\end{aligned}\label{}
\end{equation}
and this result is consistent with the field theory partition function (\ref{Z22_FT}) with $D=26$.


\section{Conclusions}\label{conclusions}

In this paper we have presented a detailed study of  twisted string theory compactified  on $S^1$. When KK and  winding states are taken into account, new and intriguing  phenomena  emerge. The spectrum of the theory  differs radically from its non-compact counterpart. Infinitely many massive higher spin fields arise.

 Interestingly enough, there are also infinitely many enhancing points at which infinitely many massless higher spin fields appear in the spectrum.  The maximal enhancement occurs at the self dual point. The algebra associated to the enhanced gauge symmetry at the self dual radius is quite intricate. However, for a particular subsector of the massless fields, the enhanced algebra was identified as  $iso(2)_L\oplus su(2)_R$.  

 On the other hand, there are points where only a small number of massless fields appears. Generically,  at such points, in addition to the usual  massless NS sector, there is a massless higher spin field. The minimal configuration occurs at $r=\sqrt{2}\pm1$ and comprises the gravity sector and a positive norm spin three field.   
 
 Perhaps the most intriguing aspect of this theory is the partition function. As in the case of the flat non-compact twisted string, $Z(\tau,\bar{\tau})$  does not converge and it leads the lack of modular invariance (\ref{nodede}).  This fact could be a signal of an inconsistency. However we found that the partition function  accurately reproduces the expected  QFT result. We checked this result up to $n=2$ and $w=2$. We believe that it holds in general for every value of the internal momentum and winding number.  Nevertheless, proving this statement could be a difficult task, since the classification of the higher spin vertex operators is unknown. In fact even in the usual string theory this classification remains an open problem, see \cite{Sasaki:1985py, Manes:1988gz} and references therein. It would be interesting to find an alternative expression of the partition function $Z$  instead of using the $Z(\tau,\bar{\tau})$. In this way we could avoid the $\tau_1$ integration and the ill-definiteness of the $Z(\tau,\bar{\tau})$.
 
 In what follows we shall present a speculation about how the modular invariance might be restored. In \cite{Lee:2017utr} and this work we have worked out string theory on ordinary spacetime with coordinates $x^{\mu}$ and  Riemann surfaces. Hence, we have used the ordinary definition of the partition function, for instance, for the bosonic part we have 
\be
Z=\int Dx [Dh_{ab}] e^{-S[x,h]}=\int \frac{d\tau d\bar{\tau}}{\tau_2} \text{Tr}\big[\text{exp}\big(2\pi\text{i}\tau_1P-2\pi\tau_2H \big)\big]\,. \label{part}
\ee
In the case of the torus worldsheet, there are two real moduli $\tau_1$ and $\tau_2$ and the integration over the worldsheet metric $h_{ab}$ translates into the integration over $\tau_1$ and $\tau_2$,  after gauge fixing. An alternative approach which is more related to the spirit of ambitwistor string \cite{Mason:2013sva} would be extending the real functional integration domain to a complex one $x\rightarrow X=X_1+i X_2$ and choose a proper integration contour \footnote{It would be equivalent to pick a section in this complex space}  \cite{Witten:2010zr} to get a finite modular partition function.  However, in this case the equivalence between the path integral and operator definition (\ref{part}) is not guaranteed.

It is not clear whether at the self dual point the infinitely many massless higher spins can be packaged in some group representation. Of course at this point either the infinite tension $\alpha^{\prime} \rightarrow 0$ or the tensionless limit $\alpha^{\prime}\rightarrow \infty$  are much more involved and a careful analysis is needed. After taking the tensionless limit we expect every  field in the theory becomes massless for any  radius. It would be worth to further explore this line and relate it with the higher spin gauge theory \cite{Didenko:2014dwa}.
On the other hand twisted ambitwistor string in torus background may be related with conformal higher spin theory, which is regarded  as a consistent interacting theory in flat space \cite{Segal:2002gd}. It  would be interesting to compare scattering amplitudes between twisted ambitwistor string theory and conformal higher spin theory \cite{Adamo:2016ple,Adamo:2018srx}. 
 
 The appearance of infinitely many negative norm states is a signal of a severe instability, even in the supersymmetric version. Providing a physical interpretation to this instability would be important and may play a role in string cosmology.   Alternatively a mechanism could exist that projects out all the negative norm states. Finding such mechanism would place twisted string theory at the same footing of the others well established string theories. It would be worth to explore in this direction.

\section*{Acknowledgement}
We thank E. Casali and Vladislav Vaganov for useful comments and suggestions.

\newpage
\appendix

\section{Circle compactification of usual string}\label{App_A}
In this section we  review the quantization of closed bosonic string
compactified on $S^1$. Let us consider closed
bosonic string compactified on $S^1$  with radius $R$. The  coordinate identification is
\begin{equation}
X^{25}= X^{25}+2\pi R\,.
\end{equation}
On the one hand univaluedness of the
vertex operator requires discrete momentum $k_{25}$
in the compact direction.
A closed string can now wind
around the compact direction, i.e.
\be
X^{25}(\sigma^0,\sigma^1)=X^{25}(\sigma^0,\sigma^1)+2\pi \omega R\label{bccs}\,.
\ee
In the case with no compact directions, the massless fields are $g$, $B$ and $\phi$. For one compact dimension, from the $(D-1)$ perspective, the metric and the B-field
give rise to two massless Kaluza-Klein  gauge fields transforming in the $U(1)_L\times U(1)_R$ group and a massless scalar.
At the self-dual radius $R=\frac{\alpha^{\prime}}{R}=\sqrt{\alpha^{\prime}}$
there are more massless states and the $U(1)_L\times U(1)_R$ gauge group  is enhanced to
$SU(2)_L\times SU(2)_R$, with six gauge fields. There are also nine scalars transforming in the bi-fundamental representation of the enhanced group.

In this case the mass-shell condition and the level matching constraint read
\be
m^2 = -k_{\mu}k^{\mu} = (k_L)^2 +\frac{4}{\alpha^{\prime}}(\bar{N}-1)= (k_R)^2 +\frac{4}{\alpha^{\prime}}(N-1) \, ,  \label{massformulaT}
\ee
or
\be
m^2 =\frac{n^2}{R^2}+\frac{\omega^2R^2}{\alpha^{\prime}}+\frac{2}{\alpha^{\prime}}(N+\bar{N}-2) \ ,
\ee
and
\be
N-\bar{N}-n\omega=0 \ ,
\ee
where
\be
k_L=\frac{n}{R} +\frac{\omega R}{\alpha^{\prime}} \qquad \qquad k_R=\frac{n}{R} -\frac{\omega R}{\alpha^{\prime}}\, .
\label{LVmatching}\ee

Let us focus now on the massless states at the self dual point. From (\ref{massformulaT}) it follows that 
\bea\label{maslessordinarycase}
k_L & = & \pm \frac{2}{\sqrt{\alpha^{\prime}}}(1-\bar{N})^{\frac{1}{2}}\ , \\  \nonumber
k_R & = & \pm \frac{2}{\sqrt{\alpha^{\prime}}}(1-N)^{\frac{1}{2}}\, .
\eea
Each of the these equations has only two solutions, $\bar{N}=0\, ,1$ and  $N=0\, ,1$. We summarize the $D$ dimensional massless states with their corresponding vertex operator in Table \ref{tablemasslessD}. 

We define the  left/right internal currents as 
\bea
J^{\pm}_{L/R}(z) & = & : \exp(\frac{\pm2i}{\sqrt{\alpha^{\prime}}}X_{L/R}(z)): \ ,
\\
J^{3}_{L/R}(z) & = &  :(\frac{i}{\sqrt{\alpha^{\prime}}})\partial X_{L/R}(z) :  \ ,
\label{current}\eea
where $a$, $\epsilon$ and $\phi$  denote the polarization of the corresponding state. For simplicity we have omitted the exponential operator $\exp(i k\cdot x(z))$ in the vertex operator.
\begin{table}[ht]
\centering
\begin{tabular}{lc}
\toprule
Level & Vertex operator
\\\midrule
$\begin{matrix}N=0\\ \bar{N}=0\end{matrix}$  \hspace{0.2cm} & $\phi_{\pm,\pm}J^{\pm}_L J^{\pm}_R$ \\ \midrule
\multirow{2}*{$\begin{matrix}N=0\\ \bar{N}=1\end{matrix}$} & $(\frac{i}{\sqrt{\alpha^{\prime}}})\bar{a}^{\pm}_{\mu}\bar{\partial}X^{\mu}  J^{\pm}_R $ \\\cmidrule{2-2}
& $\phi_{3,\pm}J^3_L  J^{\pm}_R $ 
\\ \midrule
\multirow{2}*{$\begin{matrix}N=1\\ \bar{N}=0\end{matrix}$} &  $(\frac{i}{\sqrt{\alpha^{\prime}}}) a^{\pm}_{\mu}\partial X^{\mu}  J^{\pm}_L$ \\ \cmidrule{2-2}
& $\phi_{\pm,3}  J^{\pm}_LJ^3_R $ 
\\ \midrule
\multirow{5}*{$\begin{matrix}N=1\\ \bar{N}=1\end{matrix}$} & $(\frac{i}{\sqrt{\alpha^{\prime}}})^2\epsilon_{\mu\nu}\bar{\partial} X^{\mu} \partial X^{\nu}$ \\\cmidrule{2-2}
& $(\frac{i}{\sqrt{\alpha^{\prime}}})\bar{\epsilon}_{\mu}\bar{\partial} X^{\mu} J^3_R$\\\cmidrule{2-2}
& $(\frac{i}{\sqrt{\alpha^{\prime}}})\epsilon_{\mu}\partial X^{\mu} J^3_L$ \\\cmidrule{2-2}
& $\phi_{3,3}J^3_L J^3_R$ \\ \bottomrule
\end{tabular}\label{tablemasslessD}
\caption{$25$-dimensional massless spectrum.}
\end{table}

To exhibit the $SU(2)_L\times SU(2)_R$ we should check the OPE between the currents. Here we recall the general expressions for the OPE between 
$J^{\pm}(z_1)J^{\pm}(z_2)$ and $J^{3}(z_1)J^{\pm}(z_2)$ with arbitrary momentum in the compact direction
\bea\nonumber\label{OPE}
:\exp(i k_1 X(z_1)): : \exp(i k_2X(z_2)): & = & \exp(-k_1 k_2 D(z_{12})): \exp(i k_1 X(z_1)\exp(i k_2X(z_2)): \ ,\\ 
:\partial_{z_1}X(z_1): : \exp(i k_2X(z_2)): & \sim & ik\partial_{z_1}D(z_{12}): \exp(i kX(z_2)): \ .
\eea

For ordinary string theory the two point functions correlator can be split into 
\bea
D_R(z) & = &-\frac{\alpha^{\prime}}{2}\log(z)\ , \\ \nonumber
D_L(\bar{z}) & = &-\frac{\alpha^{\prime}}{2}\log(\bar{z})\, .
\eea
Let us assume for a moment that
\be
k_1=\frac{2}{\sqrt{\alpha^{\prime}}}m \qquad\qquad k_2=\frac{2}{\sqrt{\alpha^{\prime}}}n \, .
\ee
After Taylor expanding the first line of  (\ref{OPE}) we get
\begin{equation}
\begin{aligned}
:\exp(i k_1 X(z_1):\ : \exp &(i k_2X(z_2)): \sim 
\\ 
&: \sum_{p=0}^{\infty}\frac{1}{p!}\partial_{z_2}^{(p)}\big(\exp(i k_1X(z_2))\big) \exp(i k_2X(z_2)):z_{12}^{p+2mn} \, .	
\end{aligned}\label{OPE1}
\end{equation}
Specializing to ordinary $S^1$ compactification, $(m, n) =\pm 1$. The relevant part of (\ref{OPE1}) is when $m=1$ and $n=-1$ or viceversa
\be
J^{+}(z_1)J^{-}(z_2)\sim \frac{1}{z_{12}^2}+\frac{2}{z_{12}}J^3(z_2)\, .
\ee
More general the three currents satisfy the OPE
\be
  J^{a}(z_1)J^{b}(z_2)\sim \frac{\kappa^{ab}}{z_{12}^2}+i\frac{f^{ab}{}_c}{z_{12}}J^{c}(z_2) \ ,
\ee
where $a=+,-,3$ and  $\kappa^{ab}$ and $f^{ab}{}_c$ are the Cartan metric and the structure constants of the $SU(2)$ group. The infinite dimensional algebra or Kac-Moody algebra formed by the Laurent coefficients 
\be
J^{a}(z)=\sum_{m=-\infty}^{\infty}\frac{j_m^a}{z^{m+1}} \ ,
\ee
is
\be
  \big[j^{a}_m,j^{b}_n\big]= m\delta_{m,-n} \kappa^{ab}+if^{ab}{}_cj^{c}_{m+n}\, . 
\ee
Notice that the zero modes, which are the ones that are relevant for the QFT approximation, satisfy the $su(2)$ (Lie algebra) commutation relations
\be
  \big[j^{a}_0,j^{b}_0\big]= if^{ab}{}_cj^{c}_{0}\, . 
\ee


\section{Review on the one-loop partition function with asymmetric vacuum on non-compact spacetime}\label{appB}
The full partition function including the $b$ $c$ ghosts can be computed using the usual definition of the partition function in the operator formalism,
\be
\text{Z}(\tau)=\text{Tr}\big[(-1)^F(-1)^{\overline{F}}c_0b_c\overline{c}_0\overline{b}_0\text{exp}\big(2\pi\text{i}\tau_1P-2\pi\tau_2H \big)\big]\label{partition}
\ee
\be
P={\cal L}_0-\overline{{\cal L}}_0 \qquad \qquad \qquad H={\cal L}_0+\overline{{\cal L}}_0 \,.
\ee
Where the  $F$ and $\overline{F}$ are fermionic number operators acting on the right and the left part respectively and  the ${\cal L}_0$ and  $\overline{{\cal L}}_0$ operators
\begin{equation}
\begin{aligned}
  {\cal L}_0 & = \frac{1}{2}\alpha_0^2+\hat{N}_B+\hat{N}_g-1 \ ,
  \\
  \overline{{\cal L}}_0 & = \frac{1}{2}\bar{\alpha}_0^2-\hat{\overline{N}}_B+\hat{\overline{N}}_g+1\ .	
\end{aligned}\label{}
\end{equation}

Plugging  ${\cal L}_{0}$ and $\overline{{\cal L}}_{0}$ into (\ref{partition}), after some algebra we recast the partition function as, 
\bea\label{parfunc}
\text{Z}(\tau) & = & \text{e}^{-4\pi\text{i}\tau_1}\\\nonumber
{} & {} &\text{Tr}_{R_0}\big[\text{e}^{ +\pi \text{i}\tau \alpha_0^2}\big]\text{Tr}_R\big[\text{e}^{+2\pi\text{i}\tau\hat{N}_B}\big]\text{Tr}_R\big[(-1)^{F}c_0b_0\text{e}^{+2\pi\text{i}\tau\hat{N}_g}\big]\\ \nonumber
{} & {} &\text{Tr}_{L_0}\big[\text{e}^{-\pi \text{i}\bar{\tau}\bar{\alpha}_0^2}\big] \text{Tr}_L\big[\text{e}^{-2\pi\text{i}\overline{\tau}\hat{\overline{N}}_B}\big]\text{Tr}_L\big[(-1)^{\overline{F}}\overline{c}_0\overline{b}_0\text{e}^{-2\pi\text{i}\overline{\tau}\hat{\overline{N}}_g}\big]\ . \nonumber
\eea
 For the non compact strings the zero mode physical states depend only on the combination $X_{0L}+X_{0R}$, i.e.  
\be
\text{e}^{i k\cdot X_{0L}+i k \cdot X_{0R}}=\text{e}^{i k\cdot X_{0}} \ ,
\ee 
and there is no distinction between $K_L$ and $K_R$. The base of zero mode states is given by
\be
\text{e}^{i k\cdot X_{0L}} \qquad \text{and} \qquad \text{e}^{i k \cdot X_{0R}} \ .
\ee 
After identifying $X_{0L}$ and $X_{0R}$ as $X_{0L}=X_{0R}=\frac{1}{2}X_0$ and  collecting all the pieces we get (the details of the calculation can be found in \cite{Lee:2017utr}),
\begin{equation}
\begin{aligned}
   \text{Z}(\tau)  = \int\frac{\dd^D k}{(2\pi)^D} & e^{-\pi\tau_2 \alpha^{\prime}k^2} \text{e}^{-4\pi\text{i}\tau_1} 
  \\
  &\times \Big[\sum_{N=0}^{\infty}P\big(N\big)\Big(\text{e}^{2\pi\text{i}\tau}\Big)^{N}\Big]^{(D-2)} \Big[\sum_{\overline{N}=0}^{\infty}P\big(\overline{N}\big)\Big(\text{e}^{2\pi\text{i}\overline{\tau}}\Big)^{\overline{N}}\Big]^{(D-2)}\ ,	
\end{aligned}\label{Ztau_noncompact}
\end{equation}
where $D=26$,  is the critical dimension. 
Since the second series in (\ref{Ztau_noncompact}) does not converge the partition function is not modular invariant. Instead of performing the integration on the moduli space of the torus in the fundamental domain, now we can perform the integration in the full strip $\tau_2>0$ and $| \tau_1 |<\frac{1}{2}$. Moreover we can perform first the integration in the $\tau_1$ direction. 

The relevant terms in (\ref{Ztau_noncompact}) for computing the integration in $\tau_1$ are  
\begin{equation}
\begin{aligned}
  \text{Z}(\tau) & \sim \int\frac{d^Dk}{(2\pi)^D}e^{-\pi\tau_2 \alpha^{\prime}k^2} \text{e}^{-4\pi\text{i}\tau_1}
  \\
  & \qquad\times \Big[1+P(1)\text{e}^{2\pi\text{i}\tau}+P(2)\text{e}^{4\pi\text{i}\tau}\Big]^{(D-2)} \Big[1+P(1)\text{e}^{2\pi\text{i}\bar{\tau}}+P(2)\text{e}^{4\pi\text{i}\bar{\tau}}\Big]^{(D-2)}\ .
\end{aligned}\label{}
\end{equation}
Note that terms with exponents greater than $4\pi$ will be projected out by the $\tau_1$ integration. Now using multinomial expansion 
\be
(x_1+x_2+\ldots +x_m)^n=\sum_{k_1+k_2\ldots + k_m=n}\frac{n!}{k_1! k_2!\dots k_m!}\prod_{t=1}^{m} x_t^{k_t} \ ,
\ee
with exponent $n=(D-2)$ and keeping only the relevant terms we have
\begin{equation}
\begin{aligned}
  \text{Z}(\tau) & \sim \int\frac{d^Dk}{(2\pi)^D}e^{-\pi\tau_2 \alpha^{\prime}k^2} \text{e}^{-4\pi\text{i}\tau_1}\\
{} &\qquad \times\Big[1+(D-2)P(1)\text{e}^{2\pi\text{i}\tau}+\big[(D-2)P(2)+\frac{(D-2)(D-3)}{2}P(1) \big]\text{e}^{4\pi\text{i}\tau}\Big]\\ \
{} &\qquad \times\Big[1+(D-2)P(1)\text{e}^{2\pi\text{i}\bar{\tau}}+\big[(D-2)P(2)+\frac{(D-2)(D-3)}{2}P(1) \big]\text{e}^{4\pi\text{i}\bar{\tau}}\Big]
\ ,
\end{aligned}\label{}
\end{equation}
where the integer partition are $P(1)=1$ and $P(2)=2$.
At this point it is straightforward to see that after the $\tau_1$ integration  we  get a finite contribution which is in perfect agreement with the numbers of degrees of freedom and with the expected QFT result.
\begin{equation}
\begin{aligned}
  \text{Z}(\tau_2) & = \int_{-\frac{1}{2}}^{\frac{1}{2}}\text{Z}(\tau)\dd\tau_1\\ 
{} & = \int\frac{d^Dk}{(2\pi)^D}e^{-\pi\tau_2 \alpha^{\prime}k^2} \Big(\frac{1}{2}(D-2)(D+1)\ \text{e}^{4\pi \tau_2}+(D-2)^2 +\frac{1}{2}(D-2)(D+1)\  \text{e}^{-4\pi \tau_2}\Big)\ .	
\end{aligned}\label{}
\end{equation}


\newpage

\section{Dimension of the Young tableau for the irreducible  tensor representations of $SO(D-1)$ }

Here we list the irreducible tensor representations for the little group. See \cite{MaZ} for the details. 

\begin{equation}
\begin{aligned}
  \dim \Big[~\begin{ytableau}
 {} & {} & {} & {} & {} & {}
  \end{ytableau}~\Big]	&= \frac{(D+2) (D+1) D (D-1) (D-2) (D+9)}{6!}\,,
  \\
  \dim \Big[~\begin{ytableau}
 {} & {} & {} & {} & {} 
  \end{ytableau}~\Big]	&= \frac{(D-2) (D-1) D (D+1) (D+7)}{5!} \ ,
  \\
  \dim \Bigg[~\begin{ytableau}
 {} & {} & {} & {} 
  \\
  {}
  \end{ytableau}~\Bigg]	&= \frac{4(D-3) (D-2) (D-1) (D+1) (D+5)}{5!} \,,
  \\
  \dim \Big[~\begin{ytableau}
 {} & {} & {} & {} 
  \end{ytableau}~\Big]	&= \frac{(D-2) (D-1) D (D+5)}{4!}\,,
  \\
  \dim \Bigg[~\begin{ytableau}
 {} & {} & {} \\{}
  \end{ytableau} ~\Bigg] &= \frac{3 (D-3) (D-2) D (D+3)}{4!}\,,
  \\
  \dim \Bigg[~\begin{ytableau}
 {} & {} \\ {} & {}
  \end{ytableau} ~\Bigg] &=\frac{2(D-4) (D-1) D (D+1)}{ 4!} \,,
  \\
  \dim \Big[~\begin{ytableau}
 {} & {} & {}
  \end{ytableau} ~\Big] &= \frac{(D-2)(D-1)(D+3)}{3!} \,,
  \\
  \dim \Bigg[~\begin{ytableau}
 {} & {} \\ {}
  \end{ytableau} ~\Bigg] &= \frac{2(D-3)(D-1)(D+1)}{3!} \,,
  \\
  \dim \Bigg[~\begin{ytableau}
 {} \\ {} \\{} 
  \end{ytableau} ~\Bigg] &= \frac{(D-3) (D-2) (D-1)}{3!} \,,
  \\
  \dim \Big[~\begin{ytableau}
 {} & {} 
  \end{ytableau} ~\Big] &= \frac{(D+1)(D-2)}{2} \,, 
  \\
  \dim \Big[~\begin{ytableau} {}\\{}  \end{ytableau} ~\Big] &= \frac{(D-1)D}{2} \ ,
  \\
  \dim \Big[~\begin{ytableau} {}  \end{ytableau} ~\Big] &= (D-1) \ ,
  \\
  \dim \Big[~\begin{ytableau} *(gray){}  \end{ytableau} ~\Big] &= (D-2) \ ,
  \\
  \dim \big[~\bullet~\big] &= 1 \ .
\end{aligned}\label{dimYD}
\end{equation}
%



\end{document}